\documentclass[sigconf]{acmart}

\usepackage{subcaption}
\usepackage{booktabs}
\usepackage{float}
\usepackage[draft,inline,nomargin,index]{fixme}
\fxsetup{theme=colorsig,mode=multiuser,inlineface=\itshape,envface=\itshape}
\FXRegisterAuthor{cw}{acw}{Christo}

\newcommand{\ie}{i.e.,\ }

\newcommand{\aka}{a.k.a.\ }

\newcommand{\fulltitle}{``It's not a representation of me'': Examining Accent Bias and Digital Exclusion in Synthetic AI Voice Services}
\newcommand{\fullkeywords}{Speech technology, Accent, Biases, Quantitative methods, Qualitative methods}

\copyrightyear{2025}
\acmYear{2025}
\setcopyright{cc}
\setcctype{by}
\acmConference[FAccT '25]{The 2025 ACM Conference on Fairness, Accountability, and Transparency}{June 23--26, 2025}{Athens, Greece}
\acmBooktitle{The 2025 ACM Conference on Fairness, Accountability, and Transparency (FAccT '25), June 23--26, 2025, Athens, Greece}\acmDOI{10.1145/3715275.3732018}
\acmISBN{979-8-4007-1482-5/2025/06}

\begin{document}

\makeatletter
\renewcommand{\sectionautorefname}{\S\,\@gobble}
\renewcommand{\subsectionautorefname}{\S\,\@gobble}
\renewcommand{\subsubsectionautorefname}{\S\,\@gobble}
\makeatother

\title{\fulltitle}

\definecolor{linkColor}{RGB}{6,125,233}
\hypersetup{%
  pdftitle={\fulltitle},
  pdfauthor={},
  pdfkeywords={\fullkeywords},
  bookmarksnumbered,
  citecolor=black,
  filecolor=black,
  linkcolor=black,
  urlcolor=linkColor,
  breaklinks=true,
  hypertexnames=false
}

\author{Shira Michel}
\email{michel.sh@northeastern.edu}
\affiliation{%
  \institution{Northeastern University}
  \city{Boston}
  \state{Massachusetts}
  \country{USA}
}


\author{Sufi Kaur}
\email{kaur.gur@northeastern.edu}
\affiliation{%
  \institution{Northeastern University}
  \city{Boston}
  \state{Massachusetts}
  \country{USA}
  }

\author{Sarah Elizabeth Gillespie}
\email{gillespie.s@northeastern.edu}
\affiliation{%
  \institution{Northeastern University}
  \city{Boston}
  \state{Massachusetts}
  \country{USA}
}

\author{Jeffrey Gleason}
\email{gleason.je@northeastern.edu}
\affiliation{%
 \institution{Northeastern University}
  \city{Boston}
  \state{Massachusetts}
  \country{USA}
  }
 
 \author{Christo Wilson}
 \email{cbw@ccs.neu.edu}
\affiliation{%
 \institution{Northeastern University}
  \city{Boston}
  \state{Massachusetts}
  \country{USA}
  }
  
  \author{Avijit Ghosh}
  \email{avijit@huggingface.co}
\affiliation{%
 \institution{Hugging Face and University of Connecticut}
 \city{Boston}
  \state{Massachusetts}
  \country{USA}
  }
  
  \definecolor{RED}{RGB}{255, 0, 0}
  

\renewcommand{\shortauthors}{Michel et al.}

\begin{abstract}
Recent advances in artificial intelligence (AI) speech generation and voice cloning technologies have produced naturalistic speech and accurate voice replication, yet their influence on sociotechnical systems across diverse accents and linguistic traits is not fully understood.
This study evaluates two synthetic AI voice services (Speechify and ElevenLabs) through a mixed methods approach using surveys and interviews to assess technical performance and uncover how users' lived experiences influence their perceptions of accent variations in these speech technologies.
Our findings reveal technical performance disparities across five regional, English-language accents and demonstrate how current speech generation technologies may inadvertently reinforce linguistic privilege and accent-based discrimination, potentially creating new forms of digital exclusion.
Overall, our study highlights the need for inclusive design and regulation by providing actionable insights for developers, policymakers, and organizations to ensure equitable and socially responsible AI speech technologies.
\end{abstract}



\maketitle

\section{Introduction}
Recent advances in artificial intelligence (AI) have enabled unprecedented capabilities in synthetic AI voice services.\footnote{We define \textit{synthetic AI voice services} as technologies that have both speech generation and voice cloning capabilities.} These systems can now produce highly naturalistic speech and replicate individual peoples' voices with remarkable accuracy, enabling applications across education, healthcare, and digital communications.

As synthetic AI voice services become integrated into critical systems and everyday applications, a systematic examination of their performance across diverse populations has become imperative.
Drawing on theories of linguistic capital~\cite{bbc_social_mobility, levon2022speaking} and technological embodiment~\cite{dobson2015postfeminist, gray2020intersectional, kellner2003multiculturalism, hall2024representation, nakamura2007digitizing}, we position synthetic AI voice services within broader sociotechnical systems that mediate power, identity, and social status. In particular, accents and speech patterns have historically served as mechanisms of inclusion and exclusion~\cite{levon2022speaking, matsuda1991voices, kraus2019evidence}. Systems purporting to generate synthetic AI voices, if not carefully designed and evaluated, risk amplifying existing linguistic hierarchies and accent-based discrimination~\cite{levon2022speaking,drozdzowicz2024complexities,kraus2019evidence}.

Empirical investigations of how synthetic AI voice services perform across different accents, voice characteristics, and linguistic patterns remain limited. This gap in the literature is significant given that speech patterns and accents function as fundamental markers of identity, culture, and social belonging—--factors that can profoundly influence educational opportunities, economic outcomes, and social mobility. We strive to understand the relationships between synthetic AI speech services and people's existing notions of their own accents and dialect nuances.

In this study, we employ a mixed methods approach to study speech with diverse English-language accents and pitch variations generated by two contemporary synthetic AI speech services (Speechify and ElevenLabs).
We chose these services due to their significant market presence and distinct technical approaches to speech generation and voice cloning.

Our investigation comprises three methodologically complementary components. We begin with a quantitative evaluation ($N=250$) that systematically assesses synthetic voice quality across various accent and pitch combinations. This is followed by a structured survey examining user preferences of AI generated voice clones. Our investigation concludes with in-depth, qualitative interviews ($N=26$) with native speakers from diverse linguistic backgrounds, which provide insights into how these technologies interact with personal and cultural identity.

This research aims to addresses three fundamental questions:

\begin{itemize}
    \item \textbf{RQ1:} How does synthetic voice quality vary across accent and pitch combinations for (a) stock AI voices (see \autoref{accent_prediction} and  \autoref{voice_quality}) and (b) cloned voices (see \autoref{sec:clone_AMA}) generated via these services?
    
    \item \textbf{RQ2:} If AI usage is not disclosed, do users prefer AI-generated voice clones compared to the original, recorded voices across different accent and pitch combinations (see \autoref{sec:clone_naturalness} and \autoref{sec:clone_AMA})?
    
    \item \textbf{RQ3:} How do the current capabilities and limitations of synthetic AI voice services impact users' personal, cultural, and professional experiences (see \autoref{sec:interviews})?
\end{itemize}

Our findings have broad implications for technology developers, policymakers, and organizations, by highlighting how synthetic AI voice services may inadvertently perpetuate linguistic privilege and accent-based discrimination. As these technologies become embedded in social infrastructure, their potential to reinforce existing social hierarchies and create new barriers for non-dominant accent groups underscores the urgent need for inclusive development practices and robust regulatory frameworks. This work contributes to the broader discourse on AI fairness while emphasizing the critical intersection between technological systems and cultural identity.

\section{Background and Related Work}
\label{sec:background_related_work}

We begin by discussing accent representation in the media, existing issues in speech recognition systems, and the current landscape of synthetic AI voice services.

\subsection{Accent Representation and Belongingness in Media}

Media culture plays a significant role in shaping identities and societal values by influencing how we express ourselves, communicate, and learn~\cite{cartwright2001practices}. 
These cultural frameworks reflect and reinforce understandings of gender~\cite{dobson2015postfeminist}, class, race, ethnicity~\cite{gray2020intersectional}, and other social dimensions~\cite{kellner2003multiculturalism}. 
Systems of representation operate within social contexts~\cite{hall2024representation}, shaping perceptions of morality and power dynamics~\cite{kellner2003multiculturalism}. 

Accents are a vital aspect of representation. 
In linguistics, accent refers to patterns of pronunciation, distinguishing it from dialect, which also includes grammar and vocabulary. While everyone has an accent, everyday usage often treats certain ways of speaking as ``accented'' and others as neutral. Accents are shaped by geography, class, ethnicity, and culture, and perceptions of them are socially and ideologically loaded~\cite{markl2023everyone}.

However, academics and the media have found that diverse accents are underrepresented in the media and that this has negative emotional impacts on these speakers~\cite{Rangarajan2021,mengesha2021don}. In fact, content analyses of popular media reveal how accents are often portrayed in ways that reflect and reinforce societal biases and stereotypes. 
\citet{gluszek2016does} found that accents were depicted as negative traits that hinder communication and underlie discrimination. 
Similarly, \citet{levon2022speaking} highlighted that ``Non-standard'' American and ''Foreign-Anglo'' accents were associated with less favorable perceptions of status and physical traits in American primetime television.
\citet{dobrow1998good} documented how accents were strategically used to differentiate between heroes and villains in children's animated television programs.
Even online polls ranking accent attractiveness reflect judgments that shape perceptions of speaker competence and social appeal~\cite{Bona_2015,Jarvis_2020,McLoone_2024}.

\subsection{Bias and Exclusion in Automated Speech Recognition}

Existing work highlights disparities and a lack of inclusiveness in automatic speech recognition (ASR) systems across diverse populations including race~\cite{wenzel2023can,koenecke2020racial,mengesha2021don,brewer2023envisioning}, gender~\cite{harris2024modeling, attanasio2024twists, fucci2023no}, dialect~\cite{harris2024modeling}, and speech disfluency~\cite{ada2024cultures}.
Most relevant to our work are studies that have identified limitations in ASR systems for people with non-American accents, in terms of lower comprehension accuracy~\cite{palanica2019you} and higher word error rates~\cite{kulkarni2024unveiling,tadimeti2022evaluation,ike2022inequity}. These findings motivate us to study accents in speech generation AI.


\subsection{Synthetic AI Voices}

Text-to-speech (TTS) systems support a number of different functions, such as improving reading accessibility, serving as voice assistants, streamlining customer service interactions, narrating language translations, and providing voice-overs for entertainment~\cite{Rella_2023, Rajaraman_2024}.
More recently, synthetic AI voice services have expanded into uses such as creating podcasts~\cite{NotebookLM}, producing voice-overs, generating music, and preserving voices~\cite{Burke_2022}. 
Synthetic AI voice services have also extended to different modalities, including having live two-way conversations with AI assistants \cite{chatgpt_speak, gemini_live}.

Despite these advancements, synthetic AI voice services present several concerns. 
Székely et al.~\cite{szekely2025will} argue that the socioindexical properties of AI-generated voices can influence users’ speech patterns and identity expression, potentially reinforcing normative biases by defaulting to voices that are most commonly used and culturally dominant. \citet{hutiri2024not} provide a comprehensive taxonomy of the harms associated with speech generators, emphasizing the urgent need for ethical guidelines and safeguards to mitigate these risks.

One challenge is the difficulty of distinguishing AI-generated voices from real voices. Studies show that TTS systems approach human-level quality compared to human recordings~\cite{tan2024naturalspeech} and are nearly indistinguishable from human voices~\cite{cambre2020choice}. Synthetic AI voice clones also produce speech that may be mistaken for human voices~\cite{barrington2025people}.
Realistic, AI-generated voices raise concerns about impersonation, fraud, misinformation, privacy violations, and intellectual property theft~\cite{Bethea_2024,Rose2024,Andrews_2024}.

Another challenge is the lack of specificity in how AI-generated voice systems label accents, particularly varieties of English. These accent labels are often ambiguous. The way voice service providers describe accents often does not align with linguistic definitions, and instead may reflect broader or regional classifications ~\cite{bias_voice_ai_design, lang_source}.

While prior work has focused on issues that stem from the high-accuracy of some AI-generated voice systems, our focus is on issues that may stem from low-accuracy systems, specifically with respect to the quality of accented speech produced by generative AI. This is a critical facet of synthetic speech generation, because prior work has found that people tend to be attracted to synthesized voices that exhibit characteristics---including accent---similar to their own~\cite{lee2000can,nass2001does}.
Furthermore, studies have shown that listeners receive cognitive benefits when hearing information presented in a familiar accent~\cite{njie2023talker,perry2018influences,adank2009comprehension}.
Conversely, poor accent quality could potentially contribute to poor intelligibility or understandability of synthetic speech~\cite{govender2018measuring}.
To the best of our knowledge, no prior work has investigated accent quality in AI-generated speech, or how this affects user experience and comprehension across a diverse set of people.

\section{Services}

\label{sec:services}

In this study, we primarily focus on the synthetic voice generation and voice cloning use cases in the English language. Of the various commercial providers offering these features, we investigate two synthetic AI voice services: Speechify and ElevenLabs. We chose them because they are popular---with 6\,M and 23\,M users, respectively, as of June 2024 \cite{similarweb}---and they have a diverse selection of stock AI voices and voice cloning capabilities (see \autoref{a:service_ui} for the services' user interfaces).

Launched in 2017, Speechify was initially designed to help individuals access written content. 
Since then, the platform has expanded to other purposes such as reading for leisure and entertainment, as well as increasing productivity and efficiency for work-related tasks.
Speechify has over 200 ``natural, life-like'' voices in over 60 languages and voice cloning capabilities in over 40 languages.
Speechify claims to be the ``\#1 text to speech and AI voice generator'' and they have ``the most natural, human-sounding voice overs available on the market'' \cite{Speechify}.
Speechify notes they use an autoregressive voice model for their service.

ElevenLabs debuted in 2023 with the mission to ``bridge language gaps, restore voices to those who have lost them, and make digital interactions feel more human, transforming the way we connect online''.
They emphasize their `realistic, human-like, and high-quality'' speech models in 32 languages, as well as voice cloning capabilities that purportedly capture ``human intonations and inflections''~\cite{ElevenLabs}.
ElevenLabs claims to focus on safe development, deployment, and usage of their services ~\cite{ElevenLabs_safety}.
Synthetic AI voices from ElevenLabs were generated using their \textit{Eleven Multilingual v2} model ~\cite{ElevenLabs_model}, which was the default option at the time of our study.

\section{Methodology}
\label{sec:methodology}

\begin{figure*}
    \centering
\includegraphics[width=.93\linewidth]{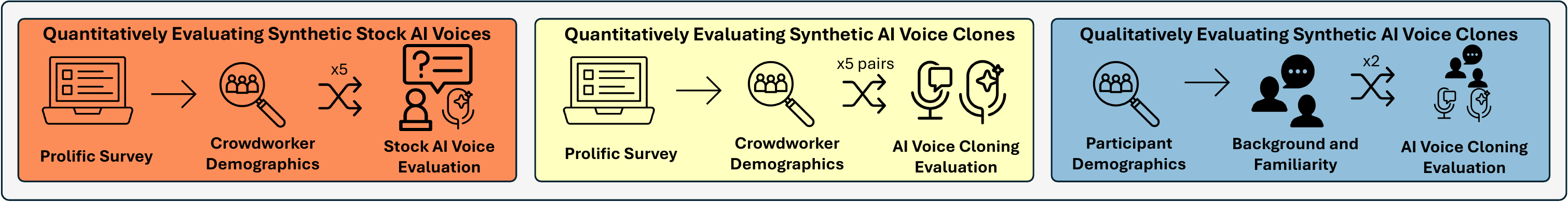}
    \caption{Overview of our mixed-methods study with surveys and interviews evaluating stock synthetic AI voices and synthetic AI voice clones. The first part involves five evaluations of synthetic stock AI voices. The second part includes evaluating five pairs of synthetic AI voices. The third part focuses on qualitative evaluation of synthetic AI voice clones in relation to participants' backgrounds and familiarity with AI.}
    \Description{A diagram summarizing the study on evaluating synthetic AI voices and synthetic AI voice clones, structured into three main parts. The first part, ``Quantitatively Evaluating Synthetic Stock AI Voices'' involves five evaluations of synthetic stock AI voices. The second part, ``Quantitatively Evaluating Synthetic AI Voice Clones'', includes evaluating five pairs of synthetic AI voices. The third part, 'Qualitatively Evaluating Synthetic AI Voice Clones,' focuses on qualitative evaluation through two sessions, incorporating participant backgrounds, familiarity with AI, and two evaluations of synthetic AI voice clones.}
    \label{fig:study_overview}
\end{figure*}

In this study, we conducted three experiments to evaluate the synthetic voice quality of English-language stock AI voices and AI voice clones\footnote{https://huggingface.co/datasets/shiramichel/ai-speech-biases} generated by Speechify and ElevenLabs. 
In addition, we interviewed native speakers to gather their perspectives on how their voices and accents were replicated by Speechify and ElevenLabs.
In this section, we present the methods that we used in our three experiments.
\autoref{fig:study_overview} presents an overview of our study procedures.
The mixed method design and procedure of our study were approved by the IRB of the authors' university. 

\subsection{Quantitatively Evaluating Stock Synthetic AI Voices}
\label{ai_voices}

\subsubsection{Text to Stock Synthetic AI Voice}

The first goal of our study was to evaluate various aspects of the synthetic AI voices that are provided as stock options in these services.
To achieve this, we generated 58 voices from English-language text with different accent and pitch combinations using the stock synthetic voice options from Speechify and ElevenLabs. 
We generated voices with high- and low-pitches. For each pitch, we generated voices with five accents: African, American, Australian, British, and Indian. We chose these parameters because (1) the pitches roughly correspond to masculine and feminine gendered voice registers, and (2) these five accents were available as options on ElevenLabs and Speechify.\footnote{As of May 2024, ElevenLabs offered only those five accents while Speechify provided these same accents as options. It is also important to note that these accent labels can be ambiguous, often encompassing a wide range of distinct regional variations.} When multiple voices were available for a given pitch and accent combination, we randomly selected three voices and generated voices from each one. For African accents, Speechify offered voices representing Kenya, Nigeria, and Tanzania, while ElevenLabs provided only Nigerian voices. Lastly, Speechify only offered one high-pitched Indian voice.

We had each AI-generated voice speak a passage randomly drawn from the CommonLit Ease of Readability (CLEAR)~\cite{crossley2021commonlit} corpus, which encompasses roughly 5,000 text excerpts with readability scores from the third to the twelfth grade-level. 
To standardize our evaluation, we selected passages at a third grade reading level to ensure that differences in voice quality were not due to linguistic complexity, but due to how voices were synthesized.

\subsubsection{Survey Study Procedure}

We recruited 307 crowdworkers from Prolific to listen to five AI-generated voices each, with accents matching their own.
We filtered to participants who spoke English, had an approval rate of at least 98\%, and 
based on their ``place of most time spent before turning 18,'' which ensured that participants would listen to accented voices that corresponded to their own accent.
Before listening, we asked participants for their self-reported age, gender, accent, and their familiarity with our five chosen accents.
After filtering out those who did not pass the attention check (see below) or whose self-reported accent was not consistent with the synthetic AI voices' accent, 57 were excluded and 250 crowdworkers remained.

Each participant listened to five voices with varied pitches generated by Speechify and ElevenLabs.
Participants were not aware that they were only listening to voices that corresponded to their own accent.
We randomized and counterbalanced the order of audio clips for each participant.
One voice served as an attention check. We anticipated that our task would take 10 minutes to complete and participants were compensated \$15 USD an hour.

\begin{figure}
    \centering
\includegraphics[width=0.5\textwidth]{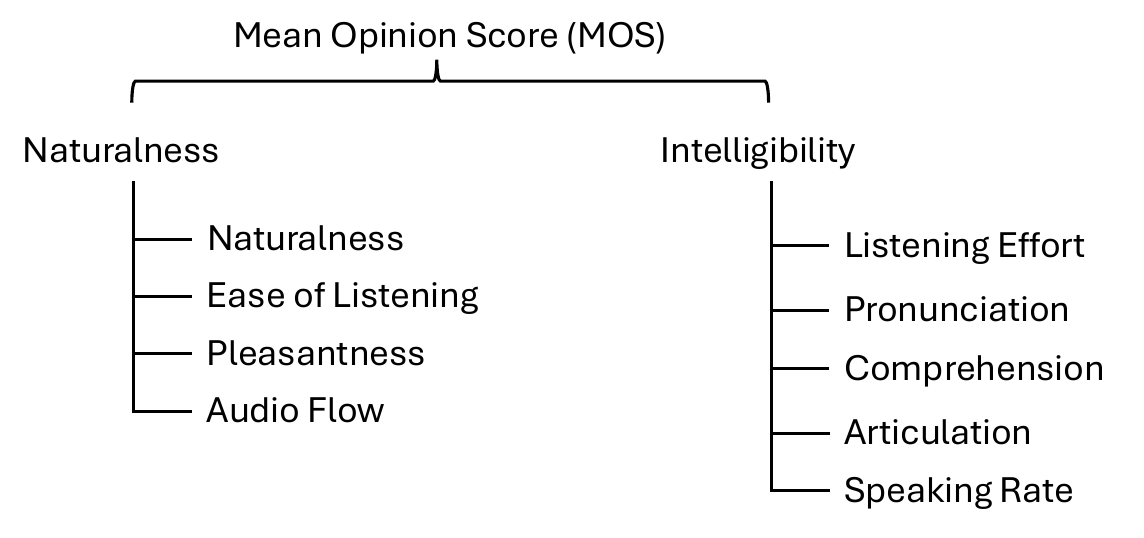}
        \caption{The two factors of the modified Mean Opinion Score (MOS) and their sub-items. Graphic is from \cite{viswanathan2005measuring}.}
        \Description{A visual of the factors and sub-items from the modified Mean Opinion Score (MOS) measure. Along the naturalness axis includes naturalness, ease of listening, pleasantness, and audio flow. Along the intelligibility axis includes listening effort, pronunciation, comprehension, articulation, and speaking rate.}
        \label{fig:MOS}
\end{figure}

We asked participants to listen to each voice for at least 15 seconds and then to label the quality of the voice using Viswanathan et al.'s modified Mean Opinion Score (MOS) measure~\cite{viswanathan2005measuring}.
MOS is a widely used~\cite{streijl2016mean} and validated measure that has been used in other studies to compare voices from TTS systems with human voices~\cite{tan2024naturalspeech,cambre2020choice}.
Viswanathan et al.'s modified MOS is a 5-point Likert scale measuring 9 attributes along the axes of naturalness and intelligibility, as shown in \autoref{fig:MOS}. 
These attributes are aggregated and averaged to produce a final MOS score.
A voice with a MOS score of 4.3--4.5 is considered to have \textit{excellent quality}~\cite{Twilio, WayWithWords_2023}.

We also asked participants to predict the accent and pitch of each voice.
Participants were asked to select one of the five predefined accents that they believed best matched the voice and if the voice was high- or low-pitched (see \autoref{a:MOS_persona_questionnaire} for the questionnaire).

\begin{figure}
        \centering
    \includegraphics[width=.3\textwidth]{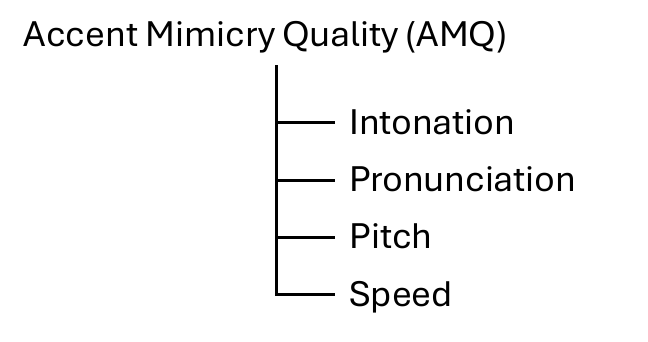}
        \caption{Our Accent Mimicry Quality (AMQ) measure for the linguistic components of speech.}
        \Description{A visual of the attributes from our Accent Mimicry Quality (AMQ) measure. These linguistic component of speech include intonation, pronunciation, pitch, and speed.}
        \label{fig:AMQ}
\end{figure}

\subsection{Quantitatively Evaluating Synthetic AI Voice Clones}
\label{quant_ai_voice_clones}

\subsubsection{Human Voice to AI Voice Clone}

The second goal of our study was to evaluate various aspects of AI-generated voice clones.
To achieve this, we generated 20 voice clones per accent and pitchusing Speechify and ElevenLabs. 
The recorded, human voices were sourced from the Speech Accent Archive~\cite{weinberger2011speech}, which is a publicly available dataset of recorded voices featuring a wide range of English speakers with many accents.
To align with our five accents of interest, we filtered the dataset based on the speakers' self-reported country of origin. 
For American, Australian, British, and Indian accents, voices were selected randomly from the respective regions.
For African accents, we randomly selected voices from Nigeria, Ethiopia, the Democratic Republic of Congo, Tanzania, and South Africa since they are some of the continent's most populous countries~\cite{Worldometer}.

\subsubsection{Survey Study Procedure}

Similar to \autoref{ai_voices}, 
we recruited 303 crowdworkers from Prolific and followed the same inclusion criteria, which resulted in 250 crowdworkers. 
In this survey, however, participants listened to five pairs of recorded and cloned voices that matched the participants' self-reported accent.
One pair served as an attention check.
As above, participants were not aware that they were only listening to voices with the same accent as them. Further, they were also not informed which clip in each pair was the recorded or the cloned voice. For each participant, the order of voice pairs was randomized (\ie Speechify or ElevenLabs first) and then the order of voices within each pair was randomized (\ie cloned voice or recorded voice first).
We anticipated that our task would take 10 minutes to complete and participants were compensated \$15 USD an hour. 

We asked participants to listen to at least 15 seconds of each voice pair and then rate (1) how natural the first voice sounded, (2) how natural the second voice sounded, and (3) how well the first voice emulated the speech style of the second voice (see \autoref{a:natural_AMA_questionnaire} for the questionnaire). 
Participants rated naturalness---\ie how closely the cloned voice resembled human speech---on a scale from 1\% to 100\%. 
Greater naturalness improves user experience by making interactions with AI systems feel more intuitive and engaging while also hiding the artificial origins of the voice \cite{cohen2004voice, kim2021designers}. 
The third task, which we referred to as Accent Mimicry Accuracy (AMA), measured how accurately the cloned voice reproduced the accent of the recorded voice. Participants evaluated AMA of the cloned voice in reproducing the linguistic components of the recorded accented voice on a scale from 1\% to 100\%.

\subsection{Qualitatively Evaluating Synthetic AI Voice Clones}
\label{qual_ai_voice_clones}

\subsubsection{Participant Recruitment}

The third goal of our study was to have people evaluate AI-generated clones of their own speech.
To achieve this, we recruited 26 participants through personal and professional networks, email outreach, and snowball sampling to participate in an interview exploring the emotional impact of hearing one's own voice reproduced by voice cloning AI (see \autoref{sec:interview-questions} for the interview questions).
Participants were required to be at least 18 years old, speak English, and self-identify as having one of the five target accents (see \autoref{table:demographics} for a summary of the participants).
The semi-structured interviews took place from August to October 2024, were about an hour in length, and participants were compensated with \$20 USD Amazon or Visa gift cards. 

\subsubsection{Interview Study Procedure}

After providing informed consent, participants submitted a one-minute voice recording of themselves reading a passage randomly selected from the CLEAR corpus. 
Using this recording, the research team created a custom voice clone for each participant using both voice cloning services.
Interviews were conducted in person and over Zoom. 
Each session began with participants being greeted and engaging in a conversation about their perceptions of their voice, their accent, and any pre-existing beliefs about AI-generated speech and voice cloning technologies. Participants also completed a demographic survey.

Participants listened first to their original recording, followed by their voice clones from each service, which were explicitly labeled.
After each listening experience, participants were asked to assess the quality of the accent mimicry in the cloned voice, referred to as Accent Mimicry Quality (AMQ).
While traditional measures like MOS assess the understandability of speech, they do not account for additional factors in the voice like accent. 
To address this, we developed AMQ as a 5-point Likert scale to evaluate how closely a cloned voice replicated the linguistic components of the original accent. 
These linguistic components of speech include intonation, pronunciation, pitch, and speed---as shown in \autoref{fig:AMQ}---which we derived from the findings of \citet{trofimovich2012disentangling}.
We validated the internal consistency of our scale using Cronbach's $\alpha$, which measures the correlation between items in the scale, achieving $\alpha=0.883$~\cite{cronbach1951coefficient}. 

In addition to assessing accent quality, participants also answered follow-up questions regarding their emotional responses to hearing their voice clones, how well the clones captured their voice and accent, and any concerns they had about voice cloning technologies. Interview participants rated how accurately each service replicated their voice using a scale from 1\% to 100\%.

\subsubsection{Interview Data Analysis}

We collected audio and video recordings of each interview and transcribed the audio via Zoom. 
After the interviews were completed, we cleaned the transcripts. 
We qualitatively analyzed the interview transcripts using a bottom-up thematic analysis. 
Authors 1 and 3 independently coded the interview responses to develop a codebook. 
They then met to discuss the codes, reconcile coding discrepancies, merge similar codes, and assign high-level themes. 
They used the digital whiteboard Miro for affinity diagramming to individually cluster the codes into larger themes. 
Following the initial grouping of the themes, they met to resolve any disagreements and to finalize what patterns were captured across the interviews.
All changes were discussed and agreed upon by both researchers. 
We did not calculate inter-rater reliability because the goal of this analysis was to identify emergent themes rather than to seek agreement, following the recommendations of \citet{mcdonald2019reliability}.

\section{Survey Results}
\label{sec:survey_results}
In this section, we present our findings from Prolific crowdworkers' evaluations of English-language stock synthetic AI voices and AI voice clones generated by Speechify and ElevenLabs. 
Participant demographics can be found in \autoref{a:prolific_demo}.
Results in \autoref{accent_prediction} and \autoref{voice_quality} are from the approach we described in \autoref{ai_voices}, while the results in \autoref{sec:clone_naturalness} and \autoref{sec:clone_AMA} are from the approach we described in \autoref{quant_ai_voice_clones}. We did not observe significant differences between high- and low-pitched voices across services, thus we present these results in \autoref{sec:pitch_analysis}.

\subsection{Stock AI Voice Accent Prediction}
\label{accent_prediction}

\begin{figure*}[t]
    \centering
    \begin{subfigure}[b]{0.4\textwidth}
        \centering        \includegraphics[height=1.3in]{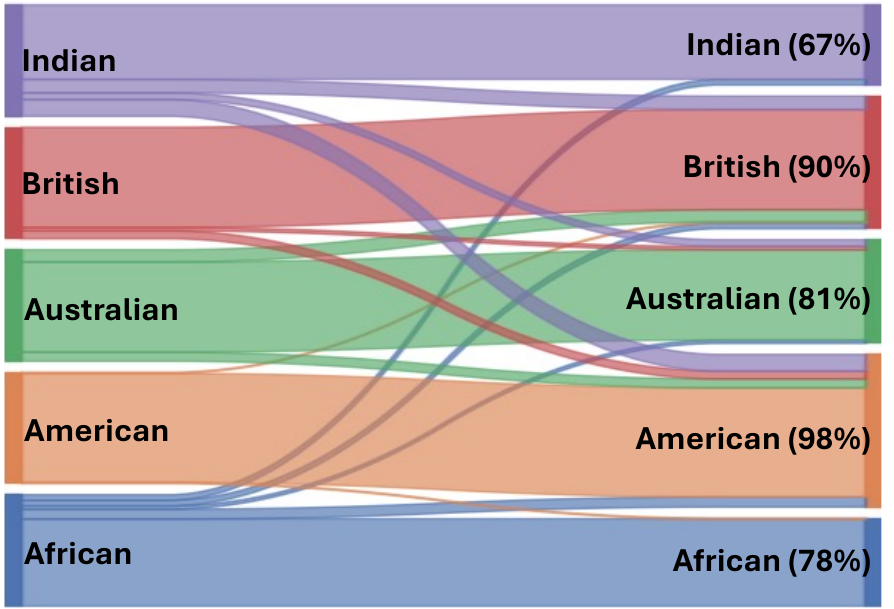}
        \caption{Speechify}
        \label{fig:sankey_predict:speech}
    \end{subfigure}%
    ~ 
    \begin{subfigure}[b]{0.4\textwidth}
        \centering
        \includegraphics[height=1.3in]{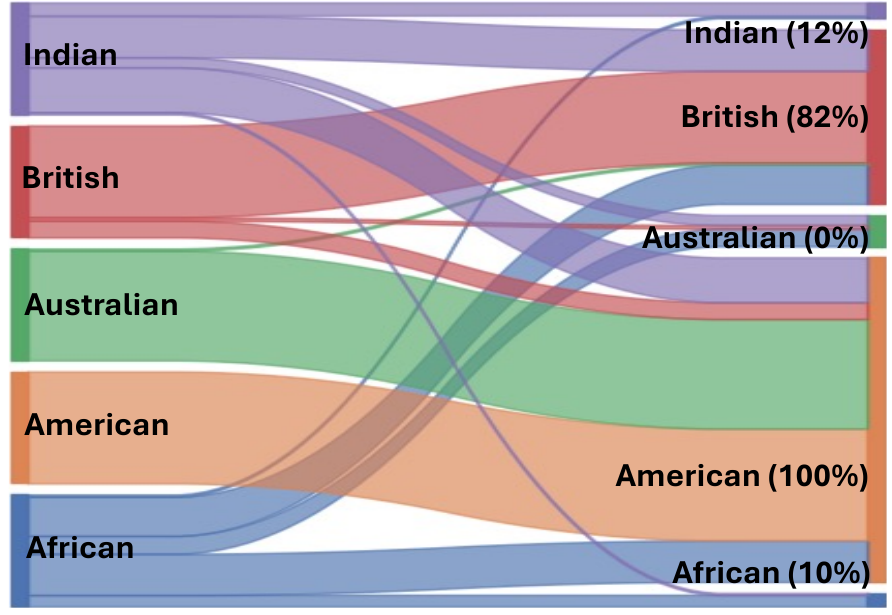}
        \caption{ElevenLabs}
        \label{fig:sankey_predict:eleven}
    \end{subfigure}
    \caption{Sankey plots showing the distribution of purported (left) and predicted (right) accents of synthetic AI voices from Speechify and ElevenLabs. Participants disagreed with the classification of voices with Indian, Australian, and African accents, especially those generated by ElevenLabs.}
    \Description{
    The left subfigure presents a Sankey plot that visualizes accent alignment between Prolific crowdworkers and Speechify's accent labels. The percentages of agreed classifications across the five accents are: Indian (67\%), British (90\%), Australian (81\%), America (98\%), and African (78\%).
    The right subfigure presents a similar Sanky plot but with accent labels from ElevenLabs: Indian (12\%), British (82\%), Australian (0\%), American (100\%), and African (10\%).}
    \label{fig:sankey_predict}
\end{figure*}

We asked Prolific crowdworkers to predict the English-language accent of the voice they were listening to without knowing they were listening to voices that matched their own accent, according to the services. 
\autoref{fig:sankey_predict} shows the mappings of the purported accents from each service to the predicted accents.
\autoref{fig:sankey_predict:speech} shows that participants agreed with the purported accents of Speechify's American and British accents (98\% and 90\% of the time, respectively). However, there was less agreement with the purported accents of Speechify's African, Australian, and Indian accents (78\%, 81\%, and 67\% of the time, respectively).
Participants listening to those accented voices interpreted them as having British or American accents.

\autoref{fig:sankey_predict:eleven} shows that participants disagreed with ElevenLabs accent designations more often than Speechify.
Like with Speechify, participants tended to agree with the purported accents of ElevenLabs' American and British accents (100\% and 82\%, respectively), but the vast majority of participants disagreed with the designations of Indian and African accented voices (only 12\% and 10\% agreed, respectively).
All participants disagreed with the designations of Australian accented voices. These findings suggest that while AI-generated American and British accents are largely perceived as intended, labels for African, Indian, and Australian voices often mismatch listener perceptions, highlighting the need for more accurate and culturally informed accent designations.

\subsection{Stock AI Voice Quality}
\label{voice_quality}

\begin{figure}
        \centering
        \includegraphics[width=\linewidth]{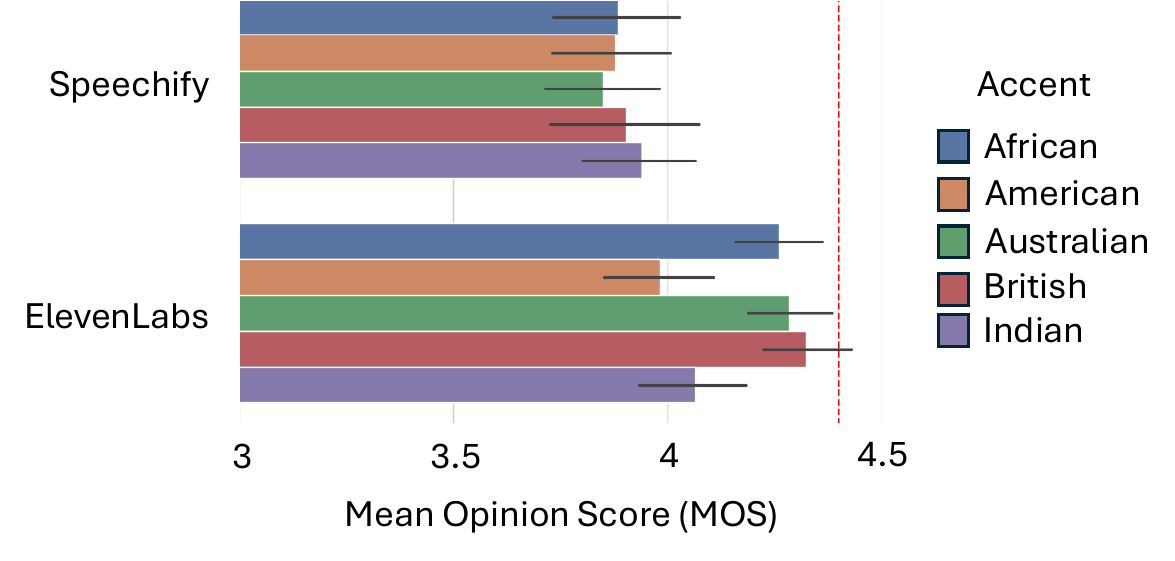}
        \caption{Perceived Mean Opinion Score (MOS) by service and accent with 95\% confidence intervals. The red line is a baseline of 4.4 which indicates \textit{excellent quality}~\cite{Twilio, WayWithWords_2023}.}
        \Description{A bar chart comparing the Mean Opinion Score (MOS) for Speechify and ElevenLabs. The x-axis represents the MOS, ranging from 3 to 4.5, while the y-axis lists the two services. Five bars represent the five accents, each with 95\% confidence intervals. A dotted red line at 4.4 marks the baseline for excellent quality, but none of the bars reach or exceed this value. }
        \label{fig:MOS_bar}
\end{figure}
\begin{figure}

        \centering
        \captionsetup{type=table}
        \small
        \begin{tabular}{l r@{}lr}
        \toprule
         & \multicolumn{2}{l}{Estimate} & (Std. Error) \\
        \midrule
        (Intercept)          & $4.23$  & $^{***}$ & $(0.13)$  \\
        African Accent       & $0.14$  &          & $(0.16)$    \\
        Australian Accent    & $0.13$  &          & $(0.16)$    \\
        British Accent       & $0.16$  &          & $(0.16)$    \\
        Indian Accent        & $0.07$  &          & $(0.16)$    \\
        Speechify            & $-0.29$ & $^{**}$ & $(0.09)$   \\
        Low-pitched Voice    & $-0.28$ & $^{**}$ & $(0.09)$  \\
        \bottomrule
        \multicolumn{4}{l}{\scriptsize{$^{***}p<0.001$; $^{**}p<0.01$; $^{*}p<0.05$}}
        \end{tabular}
        \caption{The Mean Opinion Score (MOS) regression model results. Reference levels are in comparison to ElevenLab's high-pitched, American accent.}
        \label{table:MOS_regression}
\end{figure}

We aggregated the MOS by accent and service as seen in \autoref{fig:MOS_bar}. 
A score of 4.3--4.5 indicates excellent quality~\cite{WayWithWords_2023, Twilio} and we see that no English-language accents across the two services meet this baseline. 
While Speechify produced voices with relatively consistent MOS across all accents, their ratings were less than those for ElevenLabs. However, it is important to qualify ElevenLabs' higher ratings for African, Indian, and Australian accents with the fact that almost all of our participants thought they were listening to American or British accents. For an analysis of stock AI voice quality when participants' accent labels aligned with the services' labels, see \autoref{a:MOS_correct}.

We used a linear mixed model to quantify the main effects of accent, service, and pitch on voice quality, while adjusting for random rater identity and audio clip effects. \autoref{table:MOS_regression} shows that both Speechify voices and low-pitched voices have significantly lower MOS (-0.29 and -0.28 points, respectively). We do not observe significant differences across accents. 
For an analysis of the linear mixed model with respect to each service, see \autoref{a:MOS_by_service}.

Although ElevenLabs voices received higher quality ratings than Speechify, these ratings may have been influenced by mismatches between how participants perceived the accents and how the services labeled them. This suggests that listener perceptions have a significant impact on how the quality of synthetic voices is evaluated.

\subsection{Naturalness of AI Voice Clones Across Services}
\label{sec:clone_naturalness}

\begin{figure*}[t]
        \centering
        \begin{subfigure}[t]{0.32\textwidth}
        \centering
        \includegraphics[width=1\linewidth]{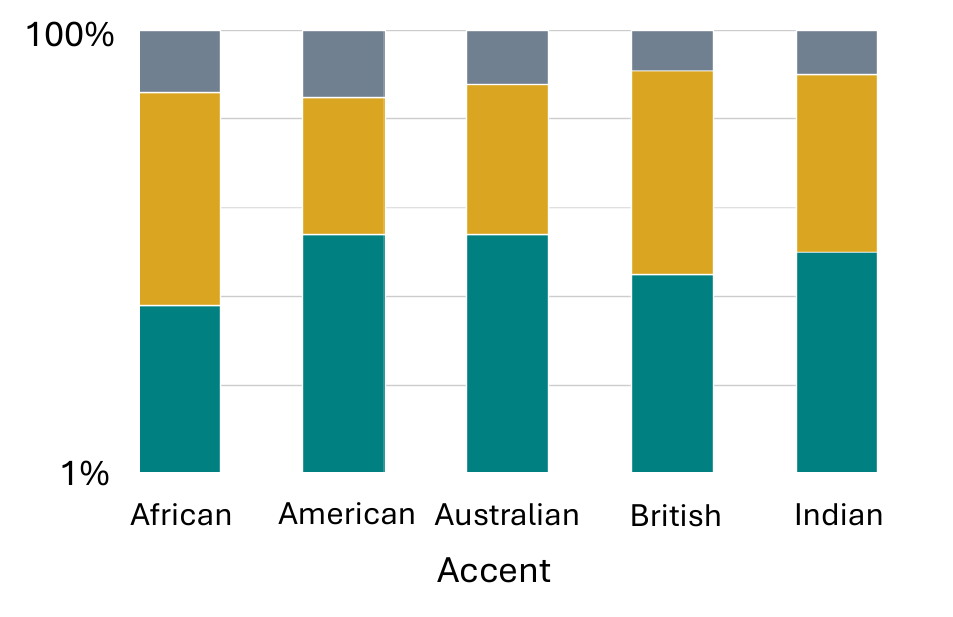}
        \caption{Speechify}
        \label{fig:interview:recreation}
    \end{subfigure}%
    ~ 
    \begin{subfigure}[t]{0.46\textwidth}
        \centering
        \includegraphics[width=1\linewidth]{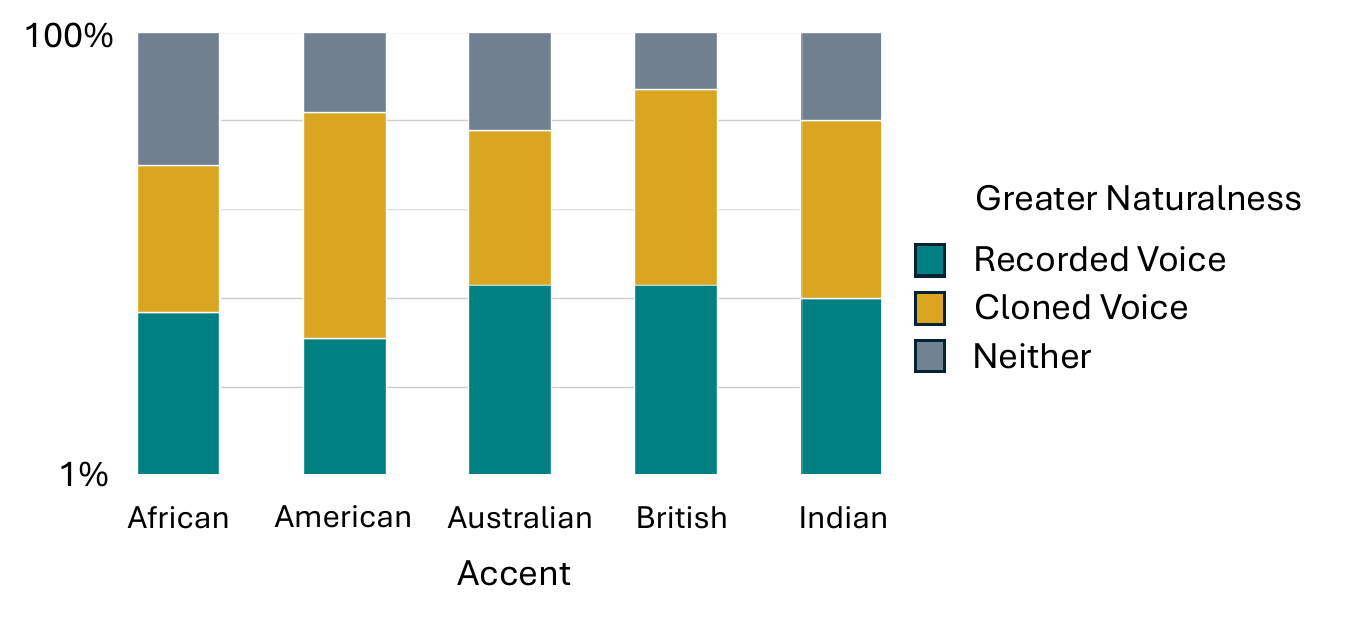}
        \caption{ElevenLabs}
        \label{fig:interview:quality}
    \end{subfigure}
    \caption{Percentage of crowdworkers evaluating greater naturalness between the recorded and cloned voice by accent and service.}
    \Description{Two stacked bar charts showing the percentage of crowdworkers' perceptions of greater naturalness between the recorded voice, voice clone voice, or neither. The left chart represents paired comparisons with Speechify across five accents, while the right chart represents comparisons with ElevenLabs. No consistent trend emerges regarding which accents are perceived as more natural across the two services.}
    \label{fig:naturalness}
\end{figure*}

We asked Prolific crowdworkers to rate the naturalness of the recorded and cloned voices on a scale from 1\% to 100\% without knowing which voices were cloned. \autoref{fig:naturalness} presents the percentage of crowdworkers that felt the recorded voice, the cloned voice, or neither had greater naturalness. 

Across both services, we observe no clear trend that one voice was perceived as having greater naturalness versus the other.
\autoref{fig:naturalness}a shows that AI-generated cloned voices with African and British accents from Speechify were deemed as more natural by participants. The ElevenLabs AI cloned voices with American and British accents were judged as sounding more natural than the recorded voices, as shown in \autoref{fig:naturalness}b. We ran a 1-way ANOVA to check whether there is a significant variation in the proportion of participants that preferred the voice clone across accents and we did not find any significant variation from either service. 
These results suggest that perceptions of naturalness in AI-cloned voices vary by accent but do not consistently favor either recorded or synthetic speech, indicating there is a more intricate connection between technical performance and user acceptance.

\subsection{Accent Mimicry Accuracy of AI Voice Clones Across Services}
\label{sec:clone_AMA}

\begin{figure}
        \centering
    \includegraphics[width=1\linewidth]{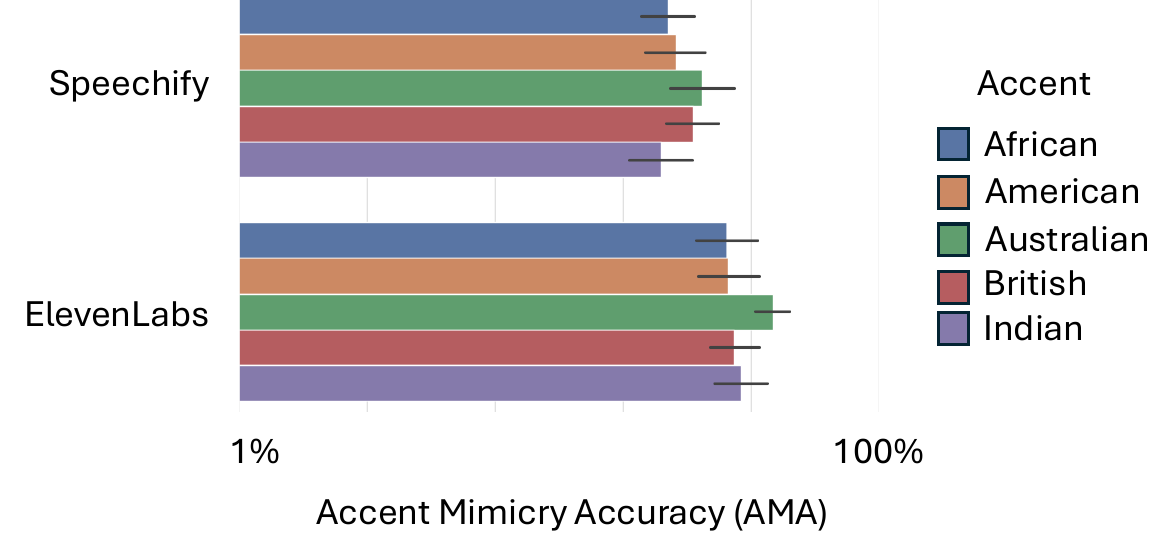}
    \caption{Perceived Accent Mimicry Accuracy (AMA) of the AI cloned voices. AMA aims to capture how closely a AI voice clone replicates the linguistic components of the original voice's accent. The bars show 95\% confidence intervals.}
    \Description{A bar chart comparing the Accent Mimicry Accuracy (AMA) for Speechify and ElevenLabs. The x-axis represents the AMA, from 1\% to 100\%, while the y-axis lists the two services. Five bars represent the five accents, each with 95\% confidence intervals. The average AMA for Speechify is lower than that of ElevenLabs, with the difference being statistically significant.}
    \label{fig:AMA}
\end{figure}

\autoref{fig:AMA} illustrates the AMA of the AI cloned voices from each service with respect to the original voices.
Both services had relatively consistent AMA across all accents, however, the AI cloned voices from ElevenLabs were judged as having slightly greater accuracy overall than the cloned voices from Speechify. The average AMA for Speechify is 69\% and for ElevenLabs is 78.5\%. These averages were significantly different from a t-test that resulted in a $t = 6.6422$ and $p < 0.05$.
The findings suggest that while both services show consistent accent mimicry, ElevenLabs voices are more often perceived as accurately capturing the original English-language accent, reflecting stronger performance in accent replication.


\section{Interview Results}
\label{sec:interviews}


In this section, we present our findings from semi-structured interviews with 26 English speakers from diverse accent backgrounds about how voice and accent fundamentally shape user experiences with AI speech technology. Before introducing them to the synthetic AI voice service experiments, we asked participants about their personal and cultural connections to their voice and accent. Their responses align with prior research, revealing that accents can lead to othering and isolation, serve as markers of identity, class, and culture, thus making code-switching both a useful skill and a burden (for more on these findings, see \autoref{sec:intimacy}). This alignment validates the representativeness of our sample.

The analysis that follows examines two key dimensions: participants' current interactions with synthetic AI voice services (\autoref{sec:familiar}), and their evaluations of voice cloning technologies, including associated concerns and implications (\autoref{sec:clone-eval}). Through this analysis, we uncover how existing system limitations and personal relationships with accent influence users' reception of emerging voice technologies.

\subsection{User Familiarity with AI Voices and AI Speech Technology}
\label{sec:familiar}

Participants demonstrated broad familiarity with synthetic AI voice applications, including voice assistants from personal devices and synthetic AI voices encountered through social media platforms. Some participants also reported encountering hurdles when using these technologies.

\subsubsection{\textbf{Voice Quality is a Spectrum from Robotic to Human-Like}}

Participants described synthetic AI voice quality on a spectrum from highly human-like to distinctly robotic. 69.23\% of responses mentioned how these systems sounded artificial. 
When describing what a realistic AI voice sounded like,
P14 described their experience nearly being deceived by a celebrity voice clone in a video, noting it sounded authentic because it replicated the celebrity's public persona.
Other participants pointed out that inconsistencies in speech were clear indicators that a voice was synthetically AI-generated.
These include mispronouncing street names (P9) and P6's full name. 

A lack of natural cadence often created dissonance when participants anticipated a human-like voice. P23 was taken aback, stating ``\textit{oh, that's not quite how I'd expect [the AI voice] to say it}.'' 
P26 characterizes detecting AI-generated speech as an ``instinctual'' process, pointing to characteristics like a slow delivery and identical intonation (P9, P23) in contrast to the speed and natural variability of human speech. 
This homogenization and lack of emotional nuance can create a sense of disconnection from the voice (P5). 
Conversely, some AI voices exhibit an overly exaggerated level of excitement, leaving listeners (P8) confused about why speech was delivered in that tone.

Participants also acknowledged the advancements AI speech technology had made in achieving natural-sounding voices.
For example, P24 described the current state of synthetic AI voices as ``developmental'' and P20 reflected how speech technology has improved from what they described as the ``classic weird'' robotic voice of earlier systems.

\begin{table*}[t]
\centering
\footnotesize
\begin{tabular}{rrrr c c c c c c rrrr}
\textbf{Participant} & \textbf{Age} & \textbf{Accent} & \textbf{Pitch} & & & & & & & \textbf{Participant} & \textbf{Age} & \textbf{Accent} & \textbf{Pitch} \\ \cmidrule{1-4} \cmidrule{11-14}
P1 & 18-29 & Indian & High & & & & & & & P14 & 30-29 & Indian & Low \\
P2 & 40-49 & British & High & & & & & & & P15 & 40-49 & Australian & High \\
P3 & 18-29 & American & Low & & & & & & & P16 & 30-39 & Australian & Low \\
P4 & 18-29 & American & High & & & & & & & P17 & 50-59 & Australian & Low \\
P5 & 18-29 & Indian & Low & & & & & & & P18 & 18-29 & African & Low \\
P6 & 18-29 & Indian & High & & & & & & & P19 & 18-29 & African & High \\
P7 & 18-29 & Indian & High & & & & & & & P20 & 18-29 & African & Low \\
P8 & 18-29 & American & High & & & & & & & P21 & 18-29 & African & Low \\
P9 & 30-39 & American & Low & & & & & & & P22 & 18-29 & African & High \\
P10 & 18-29 & Indian & Low & & & & & & & P23 & 18-29 & British & High \\
P11 & 30-39 & African & High & & & & & & & P24 & 30-39 & British & Low \\
P12 & 18-29 & American & Low & & & & & & & P25 & 18-29 & British & Low \\
P13 & 18-29 & American & High & & & & & & & P26 & 18-29 & British & Low \\
\end{tabular}
\caption{Interview participants' age, accent, and pitch.}
\label{table:demographics}
\end{table*}

\subsubsection{\textbf{American and British Accents are Overrepresented in Voice Technologies}}

Many participants (P8, P18, P19, P21, P22, P23, P26) felt that American and British accents dominate in synthetic AI voices.
Participants noted that these accents overwhelmingly define the voice standard for AI systems and excluded other accents, even saying they had never heard of an AI voice with their accent (P5, P10, P18, P21, P22, P23, P26).
Among those with non-American or non-British accents, 61.54\% of responses expressed feelings of under-representation.
Participants expressed a strong desire for AI systems to include accents that reflect their own linguistic and cultural backgrounds. 
P7 shared that they would appreciate a voice that considered their Indian accent or regional origin. 
P22 emphasized the importance of local accents, noting that most people in their country and they themselves struggle to understand the current accents available in AI systems. 
P11 highlighted the emotional impact of familiar accents, explaining ``\textit{hearing a similar voice that you are familiar with gives you a sense of belonging and a sense of familiarity as well},'' contrasting it with the distress they experience when interacting with robotic voices. 
P24 expressed that a local or country-level accent with familiar cadence and inflection would increase their trust in AI. 
P21 similarly suggested that incorporating Nigerian voices would make AI speech technologies more accessible and comprehensible for individuals in their region. 

Even with current systems offering accent options, participants felt that the voices can feel unrelatable. 
P10 shared, 
\begin{quote}
    ``\textit{But then when I heard [the voice assistant's speech] the first time, it sounded extremely typical of how, like, an Indian person speaks. I definitely don't think that's how I speak. It's not a representation of me, but I know that it's something that universally people will understand as an Indian accent, but I don't see it being very relatable to my own voice}.''
\end{quote}

This lack of representation often led to frustration. 
Participants described a range of emotions from annoyance at the homogeneity of AI accents (P24, P26) to deeper feelings of exclusion (P11, P18) and perceived mockery (P5, P10). 
For example P10 shared how they opt for the ``default voice assistant tone'' because 
\begin{quote}
    ''\textit{...sometimes it also feels like mockery, you know, like, it probably it's not the intention, but then it sometimes does feel like, you know, somebody's mocking you, like... because there is only one Indian accent that you can choose from which sounds like that and it's not relatable to you at all.}''
\end{quote}
P18 expressed resignation:
\begin{quote}
    ``\textit{Well at some point I just felt like it's a matter of not having people or not being represented enough in the society... If we had researchers who work on something like that for the African community, then it will be great. But as just having those [American and British accents]. We just know that. Well, maybe they were not meant for us. They were building them with some other people in mind}.''
\end{quote}

\subsubsection{\textbf{Understandability and Interaction Challenges}}
Participants with non-American or non-British accents reported difficulties when using AI systems that require voice for interaction. 
For example, P1 shared an experience where they attempted to set a timer, but the voice assistant mistakenly interpreted the request as a web search. 
P1 noted ``\textit{There was nobody around. There was not much noise. It should have easily picked that up, but it did not until I switched [from my Indian accent to an American accent].}'' 
Once they switched accents, the assistant successfully set the timer. 
This led P1 to question whether typing or using apps on their device would be faster and more reliable than relying on voice detection.
As P1 put it, ``\textit{How many times do [people with accents] have to try where they're like, this is a waste of time, and I'll just be better off using the application instead of using a speech assistant?}''
Similarly, P23, who was an educator, recounted how captioning software consistently failed to understand their British accent, producing incorrect transcriptions remarking, ``\textit{It came up with things I absolutely definitely did not say in a million one years.}'' 
P11, who had a similar experience with their African accent, ultimately abandoned using a voice assistant after repeated unsuccessful attempts to have their speech understood.

\begin{figure*}
    \centering
    \begin{subfigure}{0.5\textwidth}
        \centering
        \includegraphics[height=1.2in]{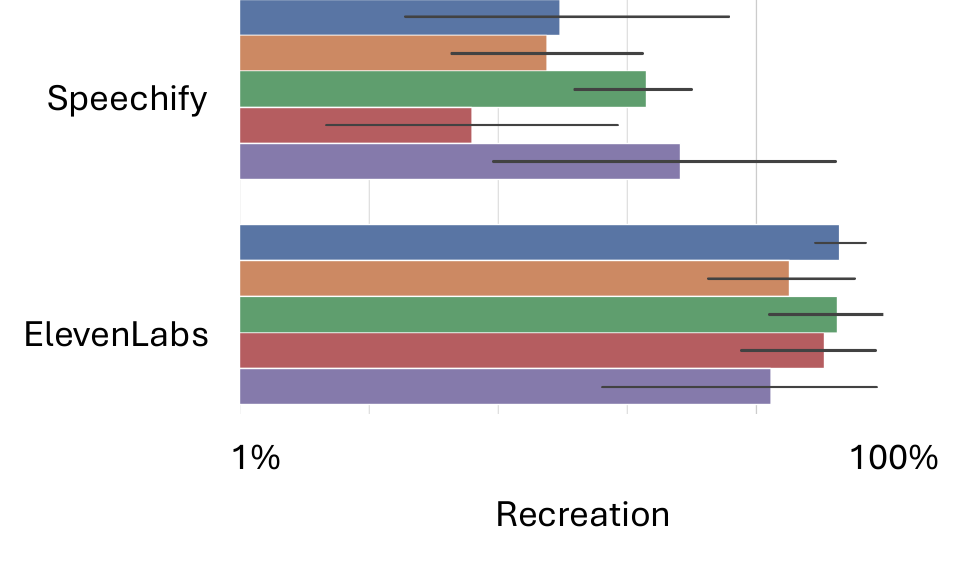}
        \caption{Voice Clone Recreation}
        \label{fig:interview:recreation}
    \end{subfigure}%
    ~ 
    \begin{subfigure}{0.5\textwidth}
        \centering
        \includegraphics[height=1.2in]{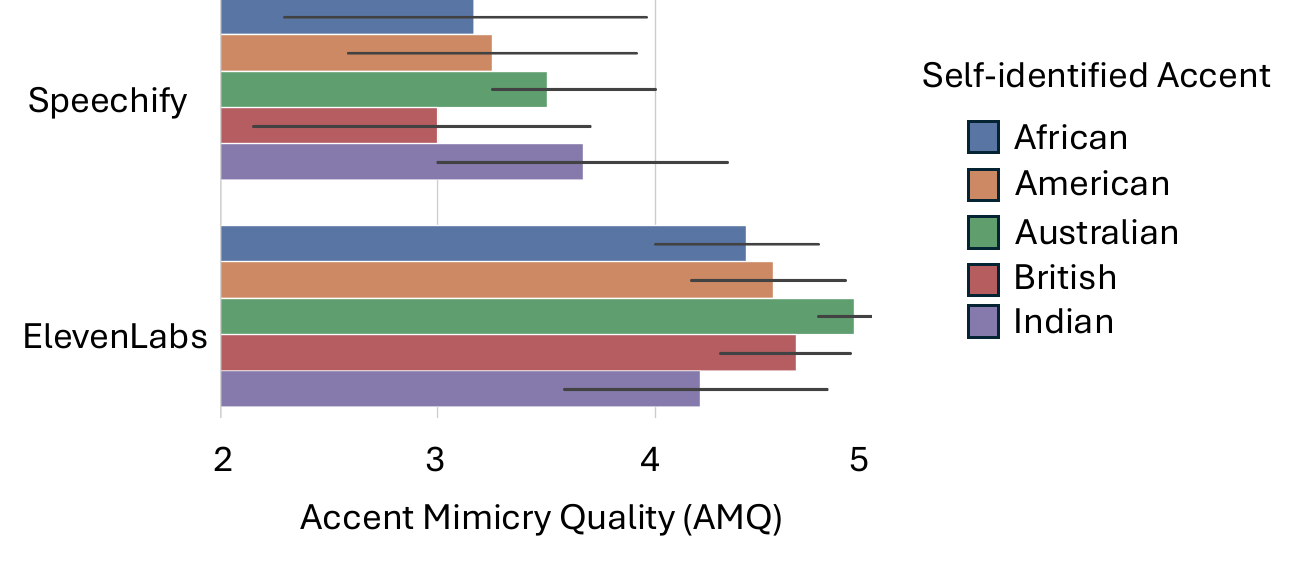}
        \caption{Voice Clone Accent Mimicry Quality (AMQ)}
        \label{fig:interview:quality}
    \end{subfigure}
    \caption{Interviewee's perceived voice clone recreation and Accent Mimicry Quality (AMQ) by service and accent. (a) ElevenLabs voice clones are observed to have greater recreations compared to the Speechify voice clones among all accents. (b) ElevenLabs voice clones also have greater perceived AMQ across all accents. The bars show 95\% confidence intervals.}
    \Description{
    The right subfigure is a bar chart comparing the percentage of voice clone recreation between Speechify and ElevenLabs. The x-axis shows the percentage of recreation, ranging from 1\% to 100\%, while the y-axis are the two services. Each of the five bars represents a different accent, accompanied by 95\% confidence intervals. ElevenLabs demonstrates higher perceived recreation percentages across all five accents compared to Speechify. The left subfigure is a bar chart comparing the Accent Mimicry Quality (AMQ) between Speechify and ElevenLabs. The x-axis represents the AMQ scores, ranging from 2 to 5, while the y-axis also lists the two services. Five bars represent the five accents, each with 95\% confidence intervals. ElevenLabs  shows higher perceived AMQ all accents compared to Speechify.
    .}
    \label{fig:interview}
\end{figure*}

\subsection{Evaluations and Implications of Voice Cloning Technology}
\label{sec:clone-eval}

Our analysis examined participants' experiences with Speechify and ElevenLabs---revealing how users evaluate these technologies' technical capabilities, cultural sensitivity, and broader implications. 

\subsubsection{\textbf{Initial Reactions}}
Reactions to the voice clones varied significantly. Participants described their reactions to Speechify as amused, curious, ``weirded out,'' scared, disappointed, or surprised. 
For example, P4 shared, ``\textit{Honestly, I'm sure you heard me laughing while I was listening to it because I was like, who is that?}'' Others were less impressed, such as P2 who critiqued, ``\textit{That was rubbish...I'm not even going to say [it was] me}''. 
Reactions to voice clones from ElevenLabs were more favorable, with participants expressing being impressed, satisfied, excited, and surprised. 
P15 remarked enthusiastically, ``\textit{Oh. wow! Wow! That wow! That's I mean, I can tell, it's different. But that's incredible. I was blown away by that.}'' 
Yet, not all feedback was positive as P23 commented, ``\textit{...it is funny how like uncomfortable it's made me feel...}''

\subsubsection{\textbf{Quality Assessment Comparisons}}

Participants generally perceived the ElevenLabs clones as more successful in recreating their voice and accent compared to Speechify across all accents (see \autoref{fig:interview:recreation}).
In particular, participants felt that their clones from Speechify had neutralized their accent (P1, P4, P10, P19, P20, P24, P25, P26), provided an incorrect accent (P2, P16, P17, P18, P23), or spoke with no accent at all (P13, P23, P24, P22). 
While fewer participants reported issues with incorrect accents from the clones generated by ElevenLabs, some participants mentioned their accent was over-emphasized or ``put through a filter'' which lead to a less authentic reproduction of their voice (P1, P3, P9, P12, P13, P26). This difference in accent recreation fidelity can be seen in \autoref{fig:interview:quality}.
 Some participants thought their clones from Speechify were more robotic due to lack of emotion (P3, P11, P19).
On the other hand, the clones from ElevenLabs were seen as more realistic because they included human characteristics like stutters, mistakes, and breaths. 
While some participants thought these imperfections were errors in the cloning process, others appreciated how realistic the voice was. P15 reflected, 
\begin{quote}
    ``\textit{[There is] humility in that you know what we create is not perfect and I think those concepts apply to this voice model as well... as a human, I'm full of imperfections, as is, you know, the other 8 billion people on the planet. [The voice clone] didn't try to iron out those imperfections and create like a homogeneous voice. It's captured my flaws as well}''. 
\end{quote}
 However, this realism also raised concerns in which eight participants remarked that the clone from ElevenLabs sounded indistinguishable from their recording. 
Participants, like P16, even wondered if this part of the study was a deception experiment, remarking ``\textit{You know where I kind of fumbled my speech. [The clone] perfectly fumbled its speech as well. It just made me think, wait a second. This feels less like a synthetic clone and more like a literal control C, control V}''.

\subsubsection{\textbf{Participant's Practical Use Cases and Concerns about the Landscape}}
In selecting the voice clone based on quality and accuracy, 66.67\% of responses preferred the clone from ElevenLabs but 52\% of responses favored a robotic AI voice clone for the ability to distinguish whether the voice was human.  
Most participants struggled to identify personal use cases for a clone generated by either service. 
These participants felt that this technology lacked relevance in their lives (P3, P11, P14, P16), expressed disinterest and aversion in using it (P2, P7), and were uncertain about positive use cases (P5, P8, P25).
However, a few participants considered using a voice clone for professional purposes, such as delivering a presentation (P1, P18, P22, P23, P26). In particular, these participants described their clones from Speechify as aspirational due to how polished, clear, and accent-diluted they sounded. 

Overall, participants expressed a range of concerns which included issues related to privacy, scams, potential misuse, lack of regulation and safeguards, misrepresentation, and broader ethical implications. These concerns align with Hutiri et al's taxonomy of consequential harms from speech generators \cite{hutiri2024not}. Some participants also voiced political concerns, particularly because the majority of interviews were conducted prior to the 2024 U.S. presidential election which heightened anxieties about potential misuses of synthetic AI voices.
Interestingly, ElevenLabs' more realistic voice clones seemed to amplify these concerns while the clones generated by Speechify generally caused fewer or no concerns. 
P2 concludes, ``\textit{I think, yeah, my takeaway would be be afraid. Be very afraid, but not of the second one [from Speechify]}''.
However, not all participants shared these concerns. 
P17 expressed confidence that worries surrounding AI speech technologies were typical of any emerging technology and would likely diminish with time, analogizing ``\textit{I'm accepting, that's the price we pay..., People who chose to stay with horse and cart were put at a disadvantage.}''

\section{Discussion}
\label{sec:discussion}
In this study, we conducted an in-depth examination of English-langauge accent bias and digital exclusion in synthetic AI voice services through a mixed-methods analysis of Speechify and ElevenLabs. Our investigation of these two prominent services reveals insights into the current state of AI voice technology. While Speechify demonstrated less accent dilution (\aka accent leveling) for non-American and non-British speakers, ElevenLabs achieved greater accuracy in capturing accent characteristics during voice cloning. However, our research shows that this accent-specific technical performance is only one dimension of how users experience and evaluate these technologies.

Our analysis presents three perspectives on synthetic AI voice services. 
First, our quantitative analysis shows that while crowdworker participants often disagreed with the classification of voices with Indian, Australian, and African accents, especially those generated by ElevenLabs, both stock AI voices and AI-cloned voices from ElevenLabs were evaluated as having higher voice and accent quality. Interview participants also generally perceived the ElevenLabs' clones as more successful in recreating their voice.
Second, our user preference data demonstrates that accent accuracy alone does not determine user satisfaction with voice clones. Naturalness ratings by crowdworkers, combined with interview participants' lack of enthusiasm for practical use cases and concerns about ethical risks, suggest a more complex relationship between technical performance and user acceptance. In particular, the high realism of ElevenLabs’ clones amplified anxieties for many. Third, our qualitative findings uncover the deeper emotional and cultural dimensions of how users interact with AI-generated versions of their own voices. Participants noted that most AI systems privilege American and British accents, and those without these accents often shared feelings of exclusion and misrepresentation.

Building on these insights, we present specific recommendations for developers, policymakers, and organizations to ensure more equitable and culturally sensitive AI speech technologies. We argue that addressing accent bias requires a multi-stakeholder approach that considers both technical performance and sociocultural impact.

\subsection{Call to Action}

\subsubsection{For Developers: Improve Technical Performance}
Our empirical findings reveal clear areas where developers must take actionable steps to create more inclusive synthetic AI voice services. The quantitative analysis of stock AI voices and voice clones demonstrated contrasting performances across services and accents, with some accents being perceived as more or less accurate. To guide improvements, we propose the Accent Mimicry Accuracy (AMA) and Accent Mimicry Quality (AMQ) measures. Interview participants emphasized that successful voice cloning must consider users' emotional connection to specific voices, their comfort with generated speech, and safety against voice clone misuse. Additionally, developers must build these speech technologies ethically~\cite{he2024cascaded}.

\subsubsection{For Policymakers: Address Misuse}
Our findings indicate an urgent need for responsible deployment and safeguards against misuse as voice technology becomes increasingly sophisticated. 
Participants in our interviews expressed greater concern over the practical challenges posed by speech technologies than about accent bias.
The human-like quality of voice recreations is a double-edged sword---while enhancing realism can improve user experience, it also risks enabling deception. This was particularly evident in participants' reactions to ElevenLabs' clones, where some initially mistook the synthetic voices for the original recordings. Notably, 52\% of responses across all accents expressed preference for more robotic voices, highlighting the importance of distinguishable AI-generated speech. To address these challenges, policymakers should prioritize provenance techniques such as watermarking for AI-generated voices and mandate other clear identification systems. While ElevenLabs currently offers a speech classifier~\cite{ElevenLabs_speech_classifier}, a broader regulatory framework is needed to prevent potential misuse across all voice synthesis applications.

\subsubsection{For Industry: Ensure Fair Treatment of Accent Diversity}
Our findings lead us to recommend that organizations adopting voice AI technologies must establish policies and practices that protect and respect accent diversity in their workplaces. Our research revealed concerning trends where American and British accents dominate commercial voice technologies, creating practical barriers to access and psychological impacts~\cite{wenzel2023can}. Many languages already face the risk of extinction~\cite{amano2014global}, and accents are experiencing what linguists term ``dialect leveling'', with speakers reducing the distinctive features of their regional accent~\cite{amano2014global}. Accent modification services market their business as making speech more ``professional'' or ``softening accents''~\cite{tomato_ai_accent_softening}. Research shows that negative perceptions of specific accents can directly impact career advancement~\cite{brusa2024voices, levon2022speaking}, highlighting why organizations must proactively prevent accent-based discrimination. Organizations that adopt voice AI technologies without considering accent diversity risk magnifying workplace inequities and contributing to the marginalization of linguistic identities~\cite{podesva2015voice}.

\subsection{Limitations and Future Work}
Our study evaluated Speechify and ElevenLabs as representative examples of current commercial AI voice technology. Although our study captured relevant user sentiments and technical capabilities using these two services, future work can extend our study by examining a broader range of synthetic voice systems. 
A major limitation of our study is that its findings are limited to English, and may not generalize to lower-resource or non-English languages.
Another limitation of our study is that the Speech Accent Archive ~\cite{weinberger2011speech} primarily consists of individuals who are immigrants, and as a result, birth country may not be a reliable descriptor of accent. We recognize that immigration can influence one's accent in ways that birth country alone does not capture. While we did not initially consider immigration status as a potential confounding factor, our decision to filter by birth country aligns with our approach to filtering Prolific crowdworkers based on “place of most time spent before turning 18” to ensure that participants listened to accented voices that are likely to be familiar with their own accent.
We also used the modified MOS measure~\cite{viswanathan2005measuring}, which does not have a validated baseline, and therefore we used the industry MOS measure's baseline. Future work could develop and validate a baseline for this modified MOS.
In our quantitative evaluation of AI voice clones, we focused on accent accuracy and naturalness metrics, but did not explicitly assess participant preferences between original and cloned voices or their ability to identify AI-generated speech. This limitation may have constrained our understanding of how users perceive authenticity in synthetic voices.

The recruitment of some interview participants through snowball sampling likely resulted in a participant group more internationally connected and culturally aware than a typical native speaker population. Although this sampling bias~\cite{amano2014global} potentially limits the generality of our interview findings, our complementary online survey helped address this limitation by providing a larger and more demographically diverse sample. Additionally, our approach of having participants read a standardized passage to ensure consistency across the interviewed participants may not have captured their natural speaking patterns or dialectical variations. This methodological choice could have influenced both the accuracy of the voice clones and participants' subsequent evaluations.

Future research could investigate several promising directions.
First, research should explore technical approaches to preserving accent authenticity while maintaining speech intelligibility, perhaps through adaptive systems that can adjust to individual user preferences and contexts. 
Second, future research could explore different interaction modalities, such as spoken conversations with AI assistants, and conduct longitudinal studies to examine how users' relationships with AI voice clones evolve over time, particularly as these technologies continue to advance in sophistication.

\section{Research Ethics and Social Impact}
\subsection{Ethical Considerations}
Our study was reviewed and approved by Northeastern University's IRB (Protocol \#2024-05-13). 
This included strategies for mitigating adverse impacts, procedures for incident reporting when necessary, and ensuring all authors completed and maintained an up-to-date CITI Program training certificate.
All crowdworkers were compensated at a rate of \$15 USD per hour, while interview participants received \$20 USD gift cards as compensation for their participation.
Participation was voluntary and there were no consequences if participants chose to withdraw, but no participant did so.
All participants provided informed consent and we encouraged questions throughout the interview study for clarity and comfort.
The collected data were only accessible to the authors.

During the interview study, we asked participants to listen to both their recorded voice and the cloned version.
We recognize that listening to an artificial clone of one's own voice can evoke emotional discomfort or distress. 
To address this in our interviews, we shared these risks with participants at the beginning of the interview and allowed participants to ask any questions. 
During interviews, our protocol was to stop the interview if any participant expressed discomfort or distress and we offered a debrief at the end of the study.
No participants reported such reactions during the study.
Once the interviews were completed and transcribed, any personally identifying information was removed.
After data was fully analyzed, participant data were deleted.
\subsection{Adverse and Unintended Impact}

Our primary adverse impact concern is unintentionally advertising these voice clone tool's potential to fuel impersonation scams. This is highlighted by P25's statement on the utility of the voice clone tools they were presented in the interview, stating \textit{``...I can definitely think of use cases, but like none that I'd participate in myself like, unless I became like a con artist...''} Secondly, despite having technical justifications to do so, we may inadvertently contribute to perpetuating overgeneralized and reductive taxonomies of regional accents, including the continent-wide category of ``African'' and subcontinent-wide category of ``Indian''. Such classifications can undermine regional diversity and may unintentionally validate problematic or colonial-era linguistic hierarchies. 
To counter this, we encourage following the \#BenderRule when naming the language or accent of study~\cite{lang_source}.
Third, by focusing exclusively on English, our study risks perpetuating the overrepresentation of English in language technologies, potentially marginalizing speakers of less-resourced languages. Finally, the study's findings, while specific to Speechify and ElevenLabs, might be extrapolated by others to represent all AI voice technologies, leading to an inaccurate understanding of the broader landscape. We strongly encourage future researchers to adapt their studies to the nuances of the specific systems that they choose to study.

\subsection{Author Positionality}
At the time of conducting this research, all authors were affiliated with academic institutions or research organizations focused on AI and machine learning, fairness, human-computer interaction, privacy, and/or security.
This study is rooted in our interest in understanding how these technologies intersect with linguistic diversity and lived experiences of users. 
We recognize how transformative the potential of synthetic AI speech generation and voice cloning technologies are as well as their broader societal implications. 
Our experiences and perspectives informs our research design, analysis, and interpretation, with the goal of advocating for advancing equitable and inclusive AI speech technologies.
The first, second, and third authors have high-pitched American accents.
The fourth and fifth authors have low-pitched American accents while the last author has a low-pitched Indian Bengali accent.

\begin{acks}
This research was supported in part by a Northwestern CASMI grant. Any opinions, findings, and conclusions or recommendations expressed in this material are those of the authors and do not necessarily reflect the views of the funders.
We would also like to thank all the participants who shared their time, insights, and experiences during the interviews, and to the anonymous reviewers for their feedback during the peer review process.
\end{acks}

\bibliographystyle{ACM-Reference-Format}
\bibliography{bibliography}

\appendix
\section{Appendix}

\subsection{Speechify and ElevenLabs}\label{a:service_ui}

\vspace{-4mm}

\begin{figure}[H]
    \centering
    \includegraphics[width=1\linewidth]{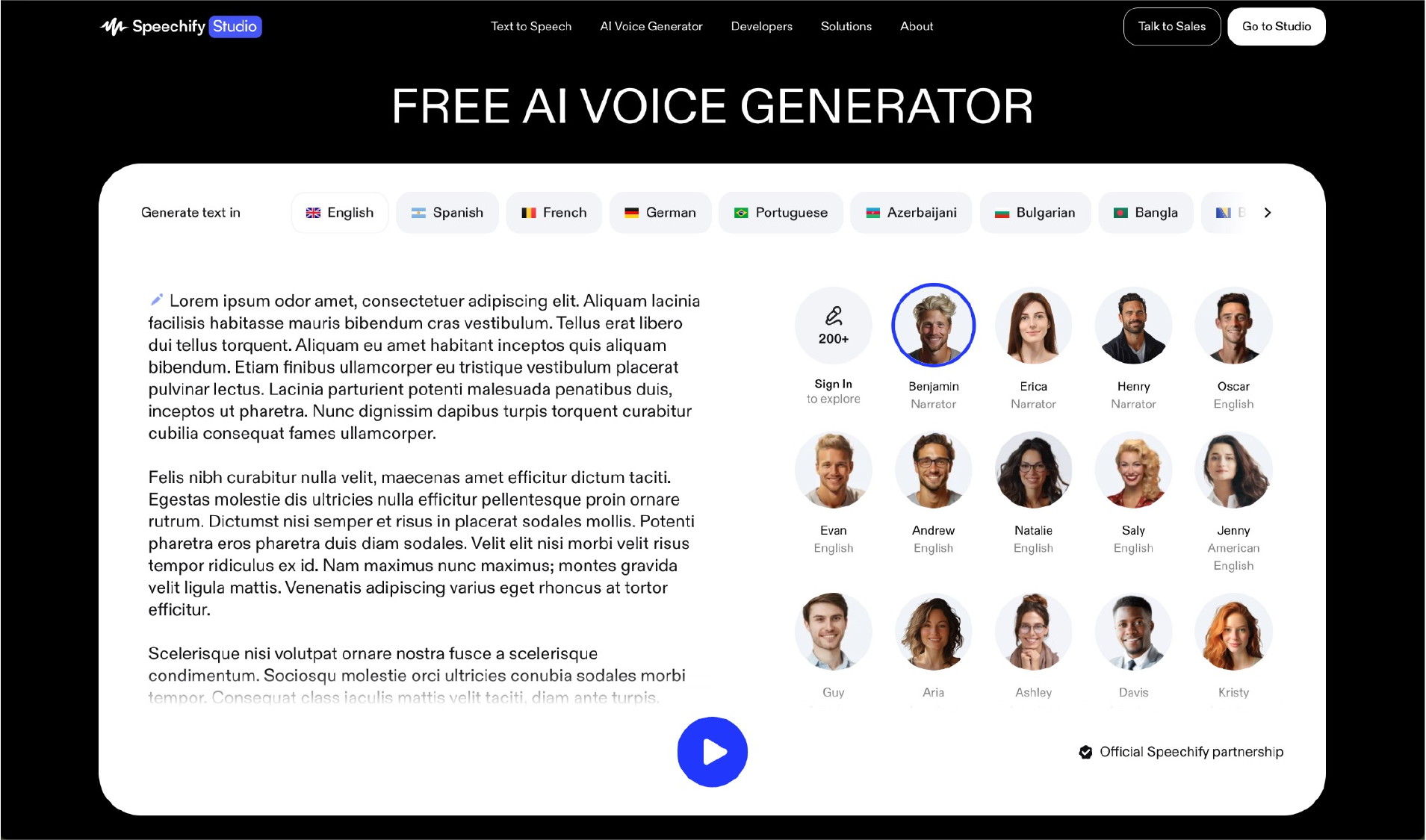}
    \caption{Speechify's user interfaces.}
    \Description{
    The screenshot is the user interface of Speechify. This screenshot includes options for the user to select what language to generated text in and options for stock synthetic AI voices.
    }
    \label{fig:speechify_ui}
\end{figure}

\begin{figure}[H]
    \centering
    \includegraphics[width=1\linewidth]{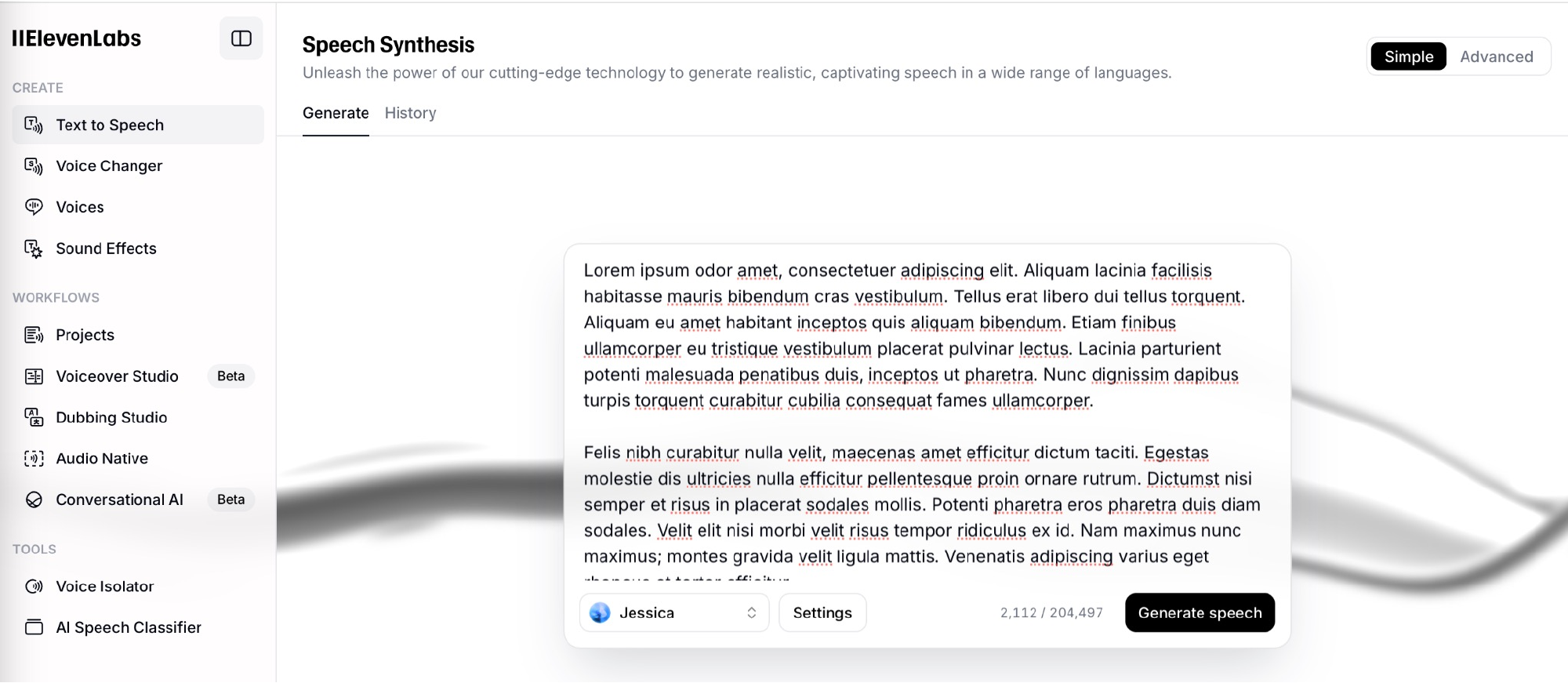}
    \caption{Speechify and ElevenLabs' user interfaces.}
    \Description{
    The screenshot is ElevenLab's user interface. This interface also has a drop down menu of what stock synthetic AI voice users can choose from as well as the side menu showcasing the services' other technical capabilities such as voice changer, voiceover studio and dubbing studio.
    }
    \label{fig:elevenlabs_ui}
\end{figure}

\subsection{Prolific Crowdworkers' Demographics}
\label{a:prolific_demo}
\vspace{-4.5mm}
\begin{figure}[H]
    \centering
    \includegraphics[height=1.6in]{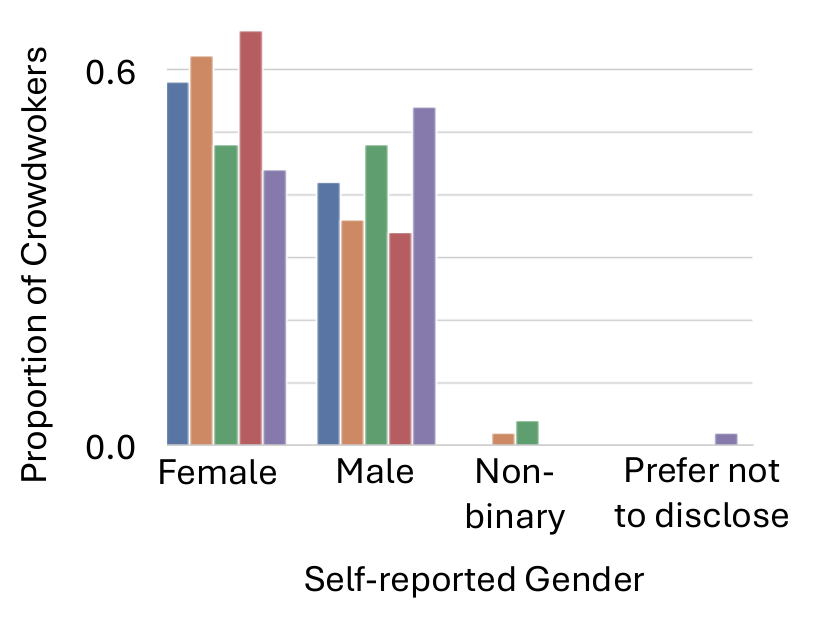}
        \caption{Prolific Crowdworkers' self-reported gender (Study 1)}
        \Description{
    The bar chart depicts the proportions of Prolific crowdworkers by their self-reported gender for Study 1. The distribution shows the majority identify as male or female.
    }
        \label{fig:demographics:gender1}
\end{figure}

\vspace{-10mm}
\begin{figure}[H]
    \centering
        \includegraphics[height=1.6in]{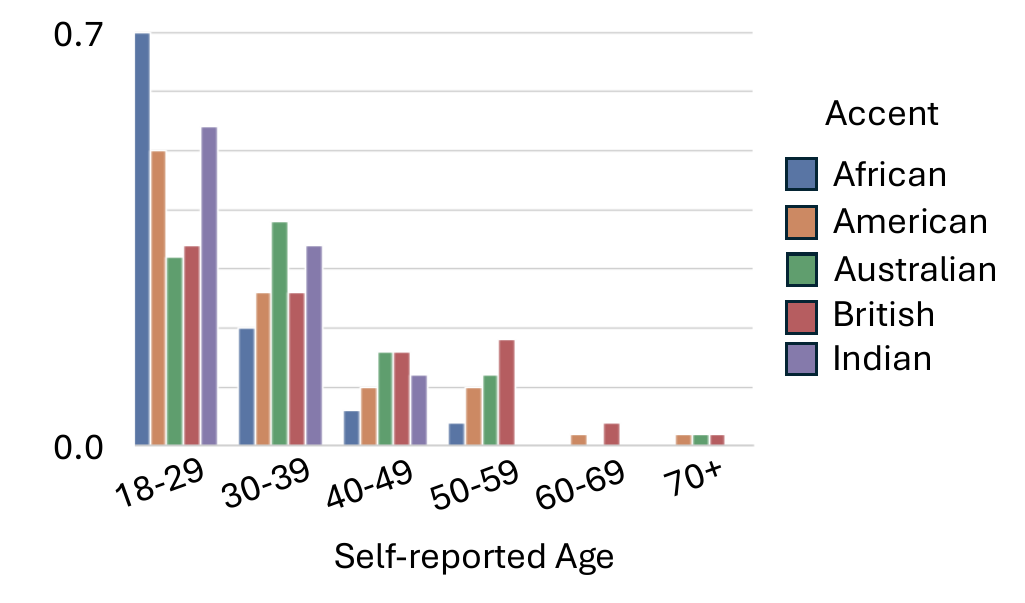}
        \caption{Prolific Crowdworkers' self-reported age (Study 1)}
        \Description{
    The bar chart illustrates the proportions of Prolific crowdworkers by their self-reported age for Study 1. Most participants are under the age of 60.
    }
        \label{fig:demographics:age1}
\end{figure}

\vspace{-0.42mm}
\begin{figure}[H]
    \centering
        \includegraphics[height=1.6in]{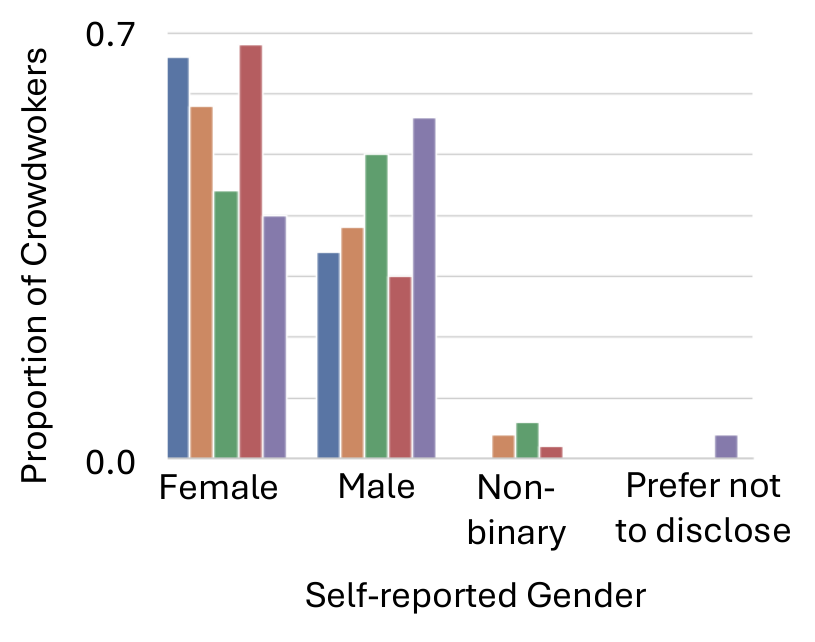}
        \caption{Prolific Crowdworkers' self-reported gender (Study 2)}
        \Description{
    The bar chart depicts the proportions of Prolific crowdworkers by their self-reported gender for Study 2. The distribution shows the majority identify as male or female.
    }
        \label{fig:demographics:gender2}
\end{figure}

\vspace{-10mm}
\begin{figure}[H]
    \centering
        \includegraphics[height=1.6in]{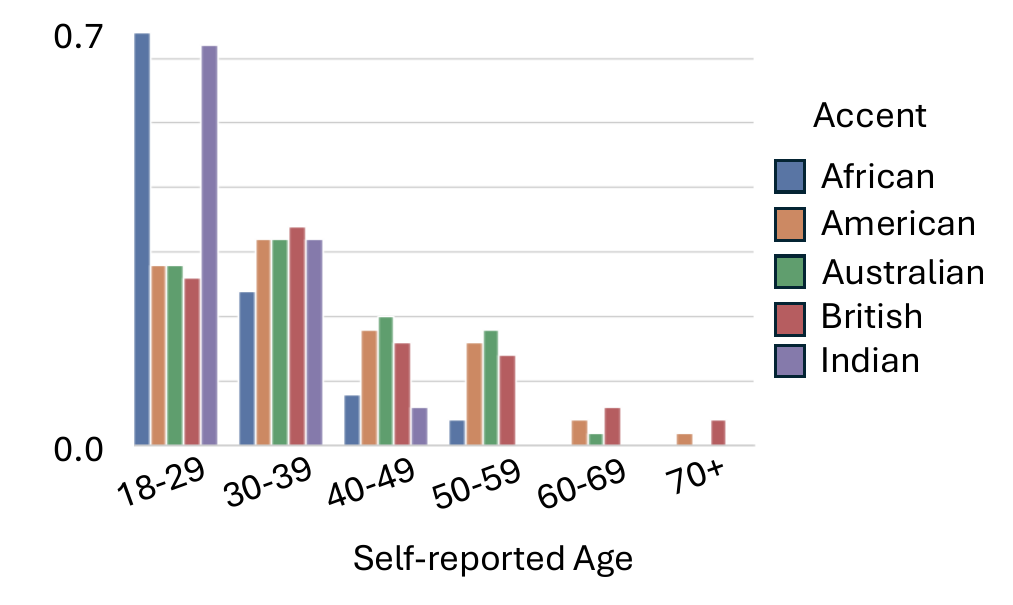}
        \caption{Prolific Crowdworkers' self-reported age (Study 2)}
        \Description{
    The bar chart illustrates the proportions of Prolific crowdworkers by their self-reported age for Study 2. Most participants are under the age of 60.
    }
        \label{fig:demographics:age2}
\end{figure}

\subsection{Modified Mean Opinion Score (MOS) and Voice Persona Questionnaire}
\label{a:MOS_persona_questionnaire}
\begin{figure}[H]
    \centering
    \includegraphics[width=0.95\linewidth]{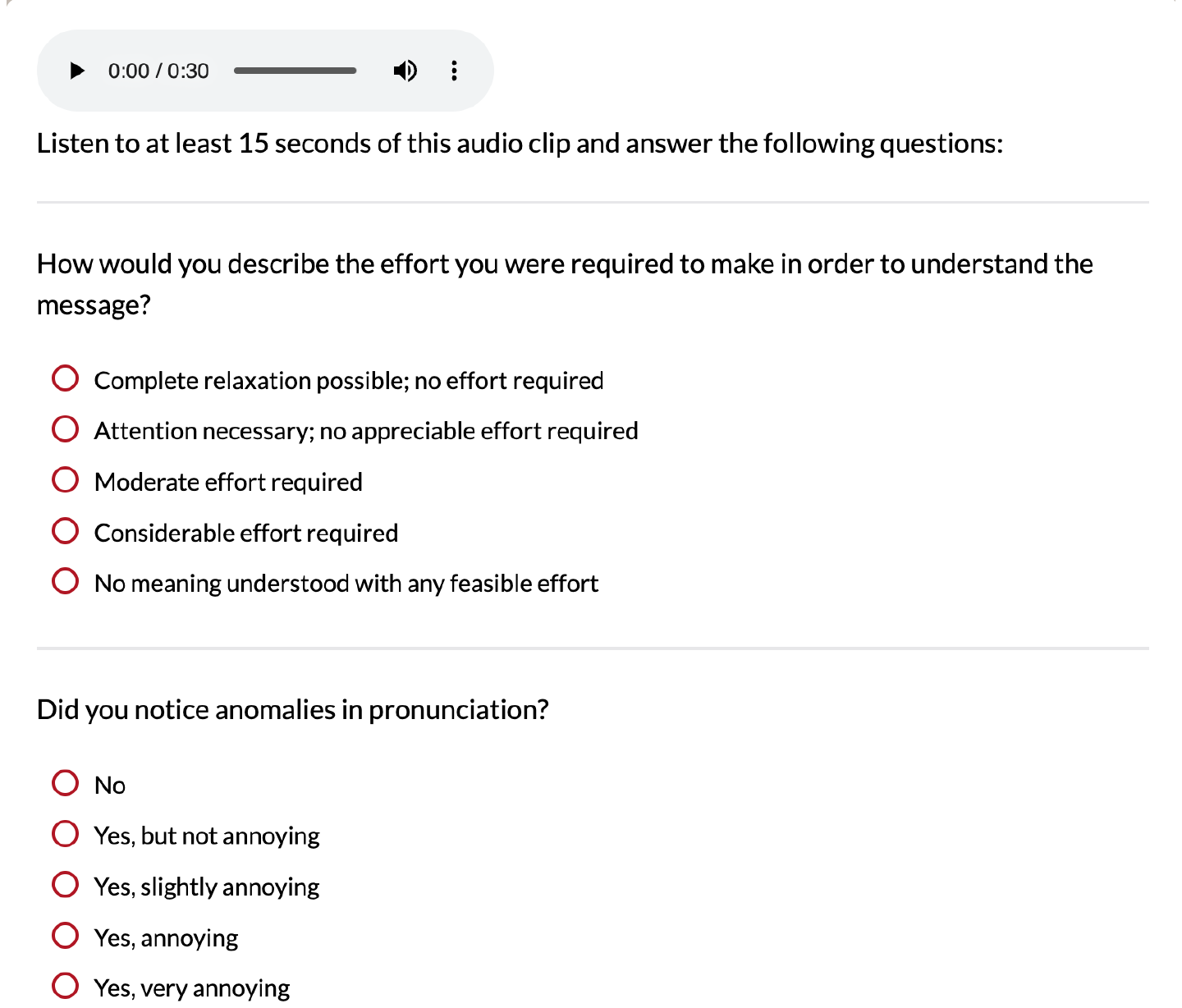}
    \label{fig:mod_MOS_1}
\end{figure}

\begin{figure}[H]
    \centering
    \includegraphics[width=0.95\linewidth]{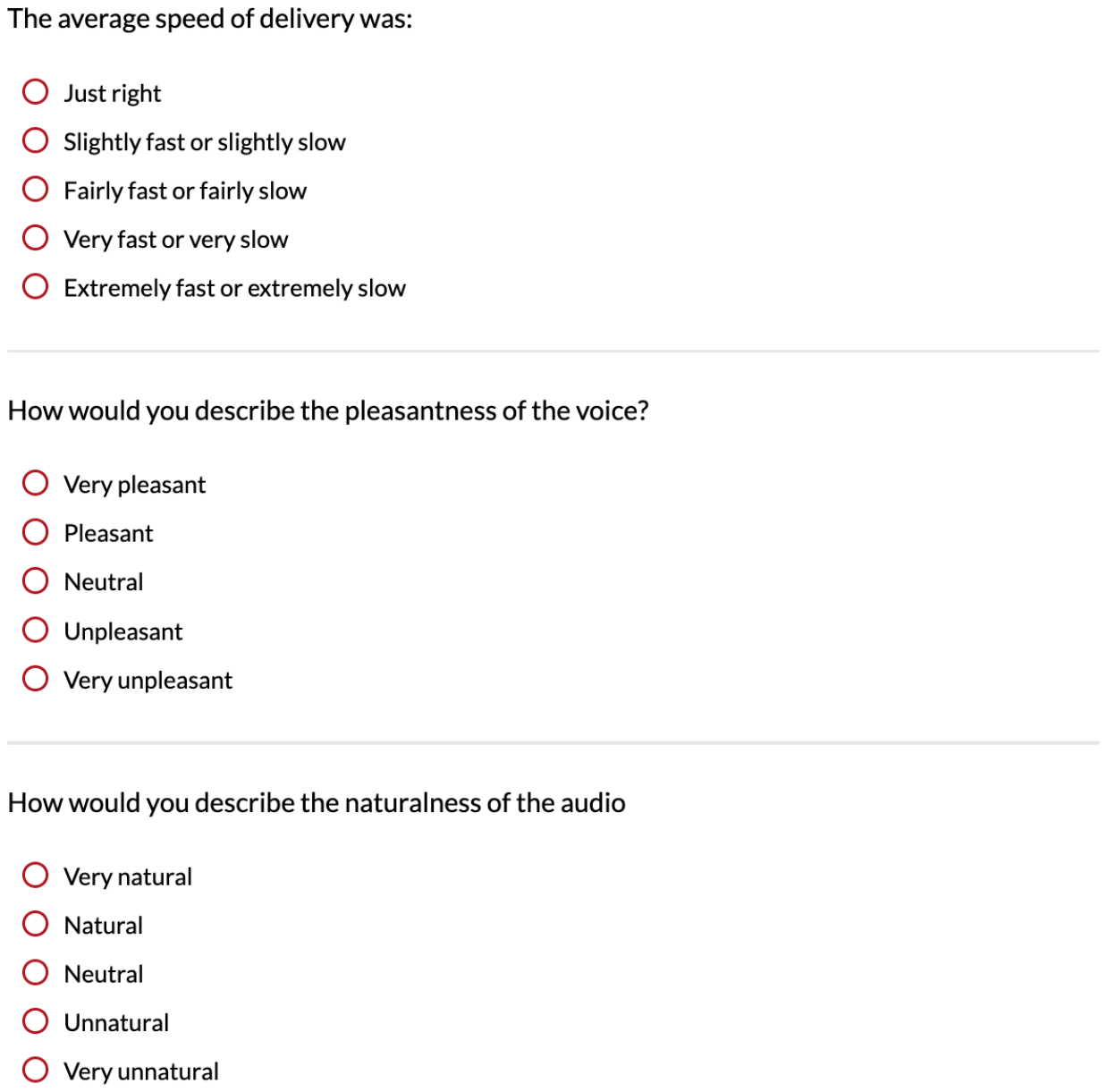}
    \label{fig:mod_MOS_2}
\end{figure}

\begin{figure}[H]
    \centering
    \includegraphics[width=0.95\linewidth]{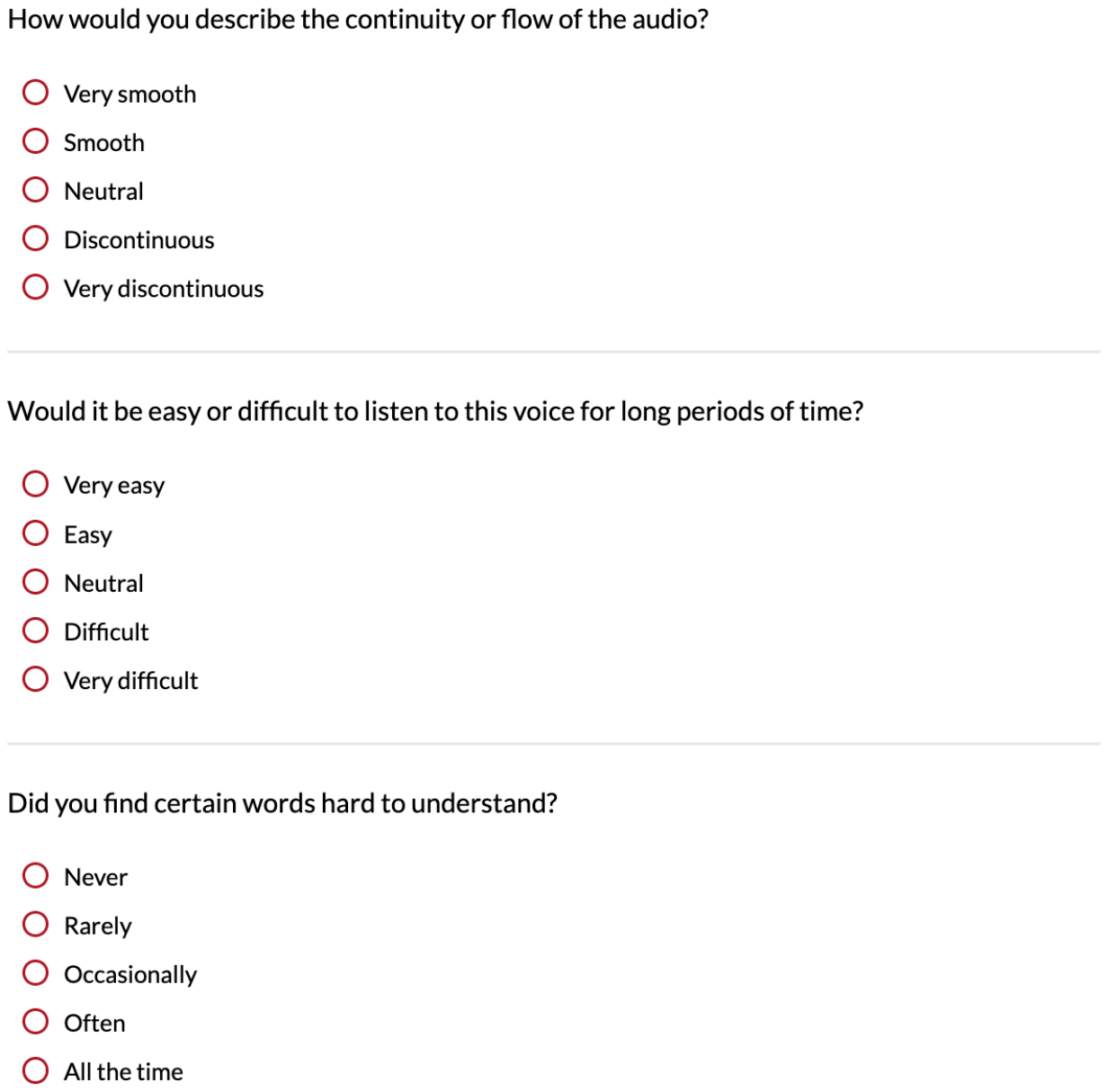}
    \label{fig:mod_MOS_3}
\end{figure}

\begin{figure}[H]
    \centering
    \includegraphics[width=0.95\linewidth]{Plots/qualtrics_study1/study1_4.pdf}
    \caption{The following is a sample survey participants completed, as described in \autoref{ai_voices}. The questionnaire instructs the participant to first listen to a voice, and answer nine questions based on the modified Mean Opinion Score (MOS) measure. The last two questions ask the participant to predict the accent and pitch persona of the voice.}
    \label{fig:mod_MOS_4}
\end{figure}

\subsection{Naturalness and Accent Mimicry Accuracy (AMA) Questionnaire}
\label{a:natural_AMA_questionnaire}
\vspace{-6mm}
\begin{figure}[H]
    \centering
    \includegraphics[width=0.95\linewidth]{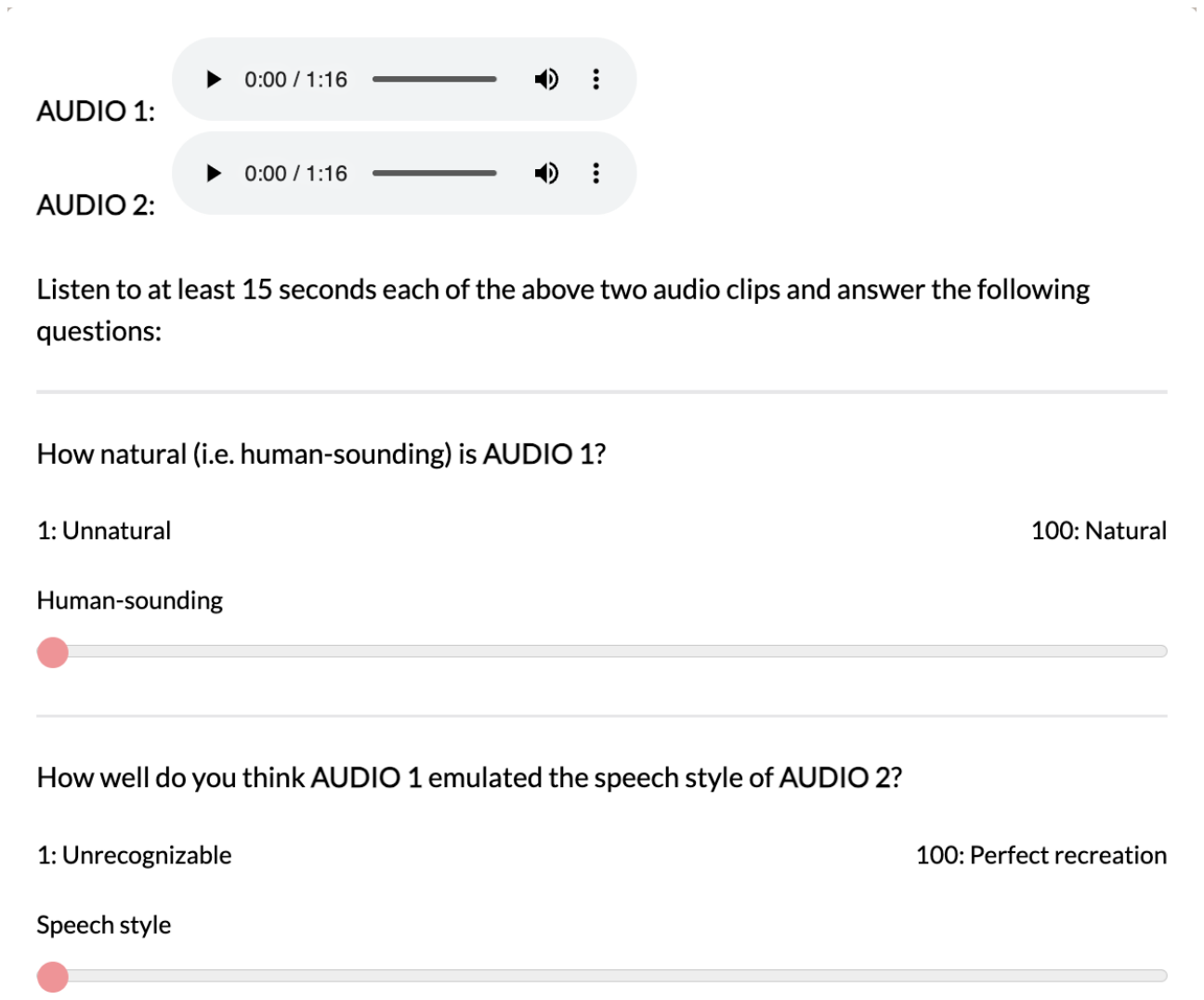}
    \label{fig:natural_ama_1}
\end{figure}%

\vspace{-40mm}
\begin{figure}[H]
    \centering
    \includegraphics[width=0.85\linewidth]{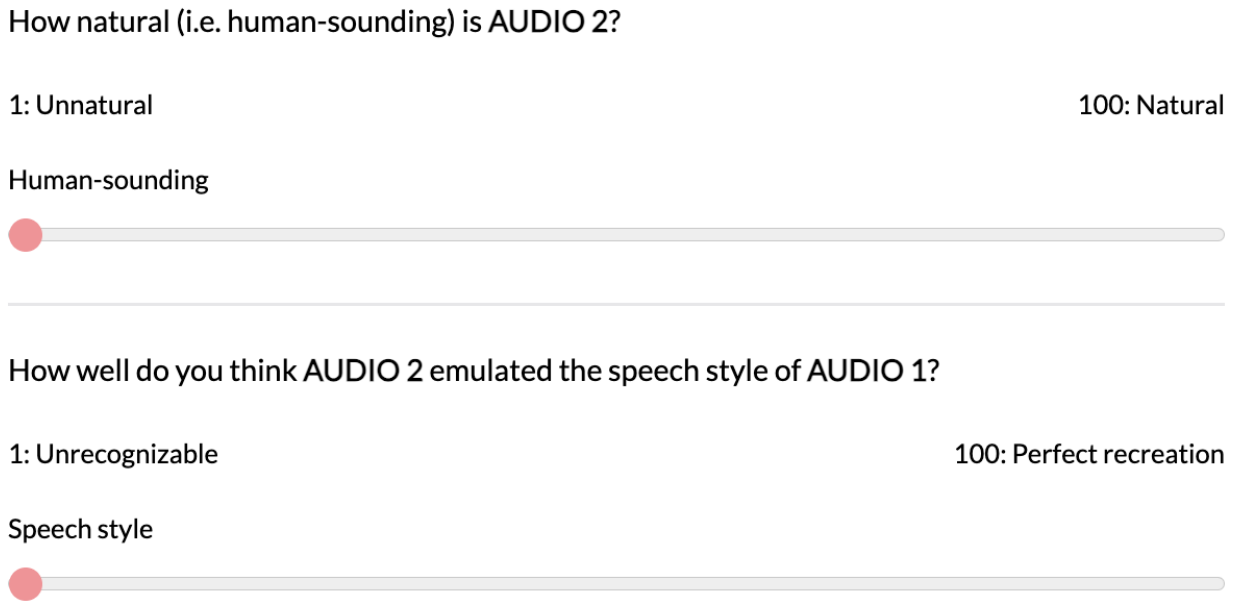}
    \label{fig:natural_ama_2}
\end{figure}

\vspace{-40mm}
\begin{figure}[H]
    \centering
    \caption{Example survey outlined in \autoref{quant_ai_voice_clones}. Participants listen to two pairs of recorded and cloned voices comparing the naturalness of each voice and how well accent mimicry is emulated.}
    \label{fig:natural_ama}
\end{figure}

\subsection{Mean Opinion Score (MOS) When Accent is Predicted Correctly}
\label{a:MOS_correct}

\begin{figure}[H]
    \centering
    \includegraphics[width=1.02\linewidth]{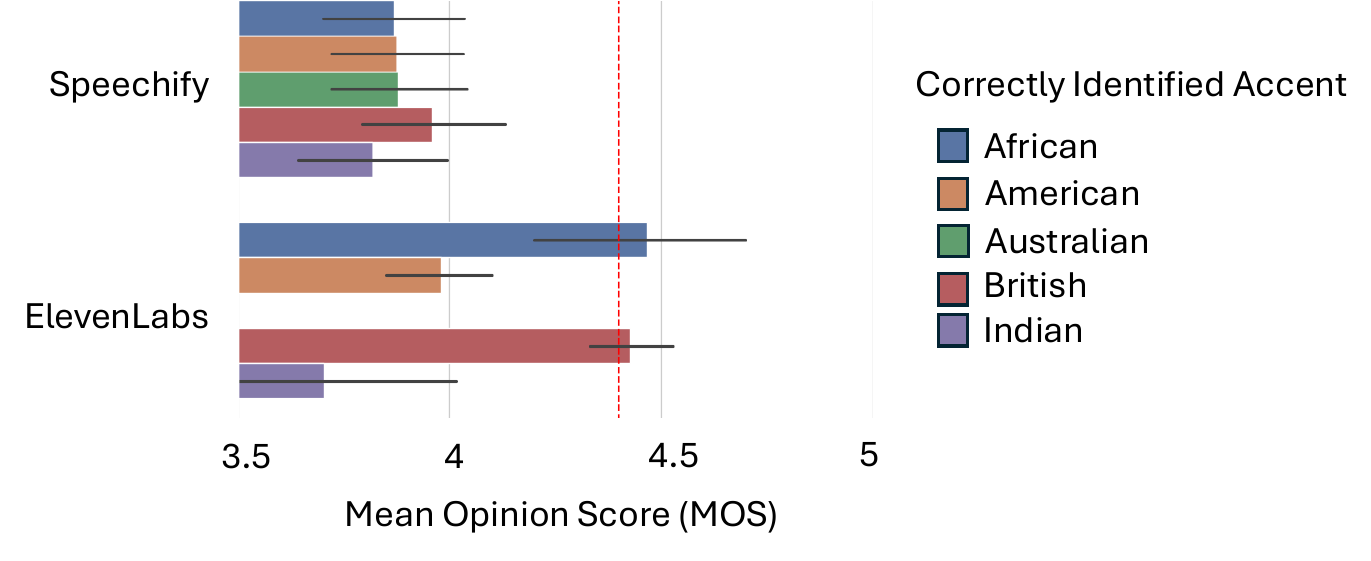}
    \caption{Perceived Mean Opinion Score (MOS) by service and accent when accent labels align with crowdworkers' classifications. The red line is a baseline of 4.4 which indicates \textit{excellent quality} for a score of 4.3--4.5~\cite{Twilio, WayWithWords_2023}. The bars show 95\% confidence intervals.}
    \Description{A bar chart comparing the Mean Opinion Score (MOS) for Speechify and ElevenLabs when accent labels align with crowdworkers' classifications. The x-axis represents the MOS, ranging from 3.5 to 5, while the y-axis lists the two services. Five bars represent the five accents, each with 95\% confidence intervals. A dotted red line at 4.4 marks the baseline for excellent quality, but none of the bars reach or exceed this value. Synthetic AI stock voices from ElevenLabs with British and African accents exceeds this threshold, indicating a higher perceived quality for these accents.}
    \label{fig:accent_correct:MOS}
\end{figure}

\begin{figure}[H]
    \centering
    \captionsetup{type=table}
    \small
    \begin{tabular}{l c}
    \hline
    & Estimate (Std. Error) \\
    \hline
    (Intercept)          & $4.22$ $(0.15)^{***}$ \\
    African Accent       & $0.15$ $(0.19)$       \\
    Australian Accent    & $0.07$ $(0.21)$       \\
    British Accent       & $0.23$ $(0.17)$       \\
    Indian Accent        & $-0.02$ $(0.21)$       \\
    Speechify            & $-0.25$ $(0.13)^{*}$  \\
    Low-pitched Voice    & $-0.31$ $(0.11)^{**}$ \\
    \hline
    \multicolumn{2}{l}{\scriptsize{$^{***}p<0.001$; $^{**}p<0.01$; $^{*}p<0.05$}}
    \end{tabular}
    \caption{The Mean Opinion Score (MOS) regression model results when accent was predicted correctly. Reference levels are in comparison to the American Accent,
    ElevenLabs, and high-pitched voices.}
    \label{table:MOS_regression_correct}
\end{figure}

\subsection{Mean Opinion Score (MOS) Regressions by Service}
\label{a:MOS_by_service}

\vspace{-2mm}
\begin{figure}[H]
    \centering
    \captionsetup{type=table}
    \small
    \begin{tabular}{l c}
    \hline
    & Estimate (Std. Error) \\
    \hline
    (Intercept)          & $4.09$ $(0.20)^{***}$ \\
    African Accent       & $0.02$ $(0.26)$       \\
    Australian Accent    & $-0.04$ $(0.26)$       \\
    British Accent       & $-0.004$ $(0.26)$       \\
    Indian Accent        & $0.15$ $(0.29)$       \\
    Low-pitched Voice    & $-0.39$ $(0.16)^{*}$ \\
    \hline
    \multicolumn{2}{l}{\scriptsize{$^{***}p<0.001$; $^{**}p<0.01$; $^{*}p<0.05$}}
    \end{tabular}
    \caption{The Mean Opinion Score (MOS) regression model results from Speechify. Reference levels are in comparison to the American Accent and high-pitched voices.}
    \label{table:MOS_regression_Speechify}
\end{figure}

\vspace{-10mm}
\begin{figure}[H]
    \centering
    \captionsetup{type=table}
    \small
    \begin{tabular}{l c}
    \hline
    & Estimate (Std. Error) \\
    \hline
    (Intercept)          & $4.09$ $(0.10)^{***}$ \\
    African Accent       & $0.28$ $(0.14)^{*}$       \\
    Australian Accent    & $0.31$ $(0.14)^{*}$      \\
    British Accent       & $0.33$ $(0.14)^{*}$       \\
    Indian Accent        & $0.07$ $(0.14)$       \\
    Low-pitched Voice    & $-0.21$ $(0.08)^{*}$ \\
    \hline
    \multicolumn{2}{l}{\scriptsize{$^{***}p<0.001$; $^{**}p<0.01$; $^{*}p<0.05$}}
    \end{tabular}
    \caption{The Mean Opinion Score (MOS) regression model results from ElevenLabs. Reference levels are in comparison to the American Accent and high-pitched voices.}
    \label{table:MOS_regression_ElevenLabs}
\end{figure}

\subsection{Voice Quality Analysis Across Pitch and Service}
\label{sec:pitch_analysis}
\begin{figure}[H]
    \centering
    \begin{subfigure}[t]{0.25\textwidth}
        \centering
        \includegraphics[width=0.7\linewidth]{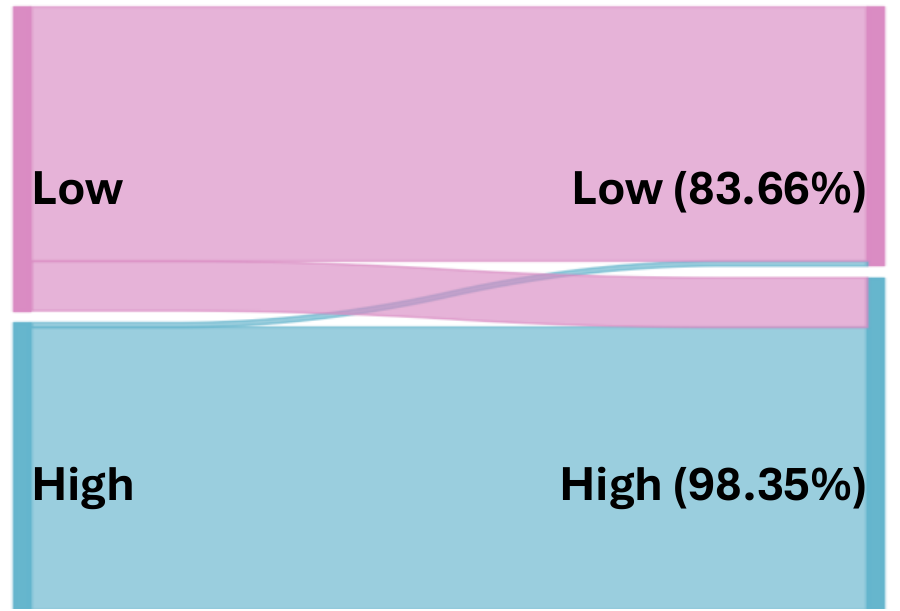}
        \caption{Speechify}
        \label{fig:pitch:speechify}
    \end{subfigure}%
    ~ 
    \begin{subfigure}[t]{0.25\textwidth}
        \centering
        \includegraphics[width=0.7\linewidth]{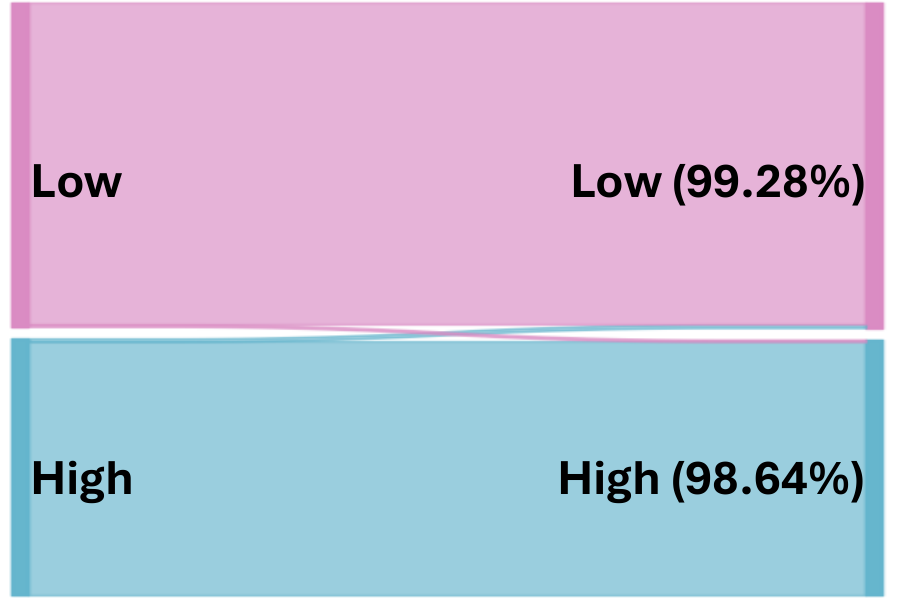}
        \caption{ElevenLabs}
        \label{fig:pitch:MOS}
    \end{subfigure}
    \caption{Sankey plots showing the distribution of purported (left) and predicted (right) pitch of synthetic AI voices from Speechify and ElevenLabs. Some participants felt that purportedly low-pitched voices from ElevenLabs were actually high-pitched.}
    \Description{The left subfigure presents a Sankey plot that visualizes pitch alignment between Prolific crowdworkers and Speechify's pitched voice labels. The percentages of agreed classifications across the two accents are: Low (99.28\%) and High (99.64\%).
    The right subfigure presents a similar Sanky plot but with pitch labels from ElevenLabs: Low (83.66\%) and High (98.35\%).}
    \label{fig:pitch}
\end{figure}

\begin{figure}[H]
    \centering
    \includegraphics[width=0.9\linewidth]{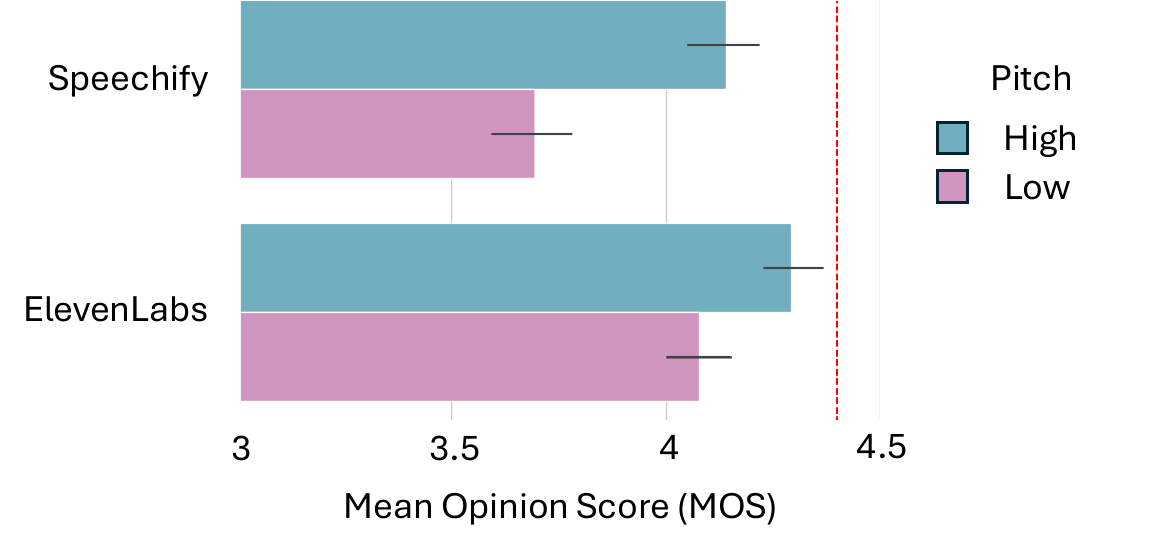}
    \caption{Perceived Mean Opinion Score (MOS) by service and pitch. The red line is a baseline of 4.4 which indicates \textit{excellent quality} for a score of 4.3--4.5~\cite{Twilio, WayWithWords_2023}. The bars show 95\% confidence intervals.}
    \Description{A bar chart comparing the Mean Opinion Score (MOS) for Speechify and ElevenLabs. The x-axis represents the MOS, ranging from 3 to 4.5, while the y-axis lists the two services. Two bars represent the two pitched voices, each with 95\% confidence intervals. A dotted red line at 4.4 marks the baseline for excellent quality, but none of the bars reach or exceed this value.}
    \label{fig:pitch:MOS}
\end{figure}

\subsection{Interview Questions}
\label{sec:interview-questions}
\begin{figure}[H]
    \centering
    \includegraphics[width=0.9\linewidth]{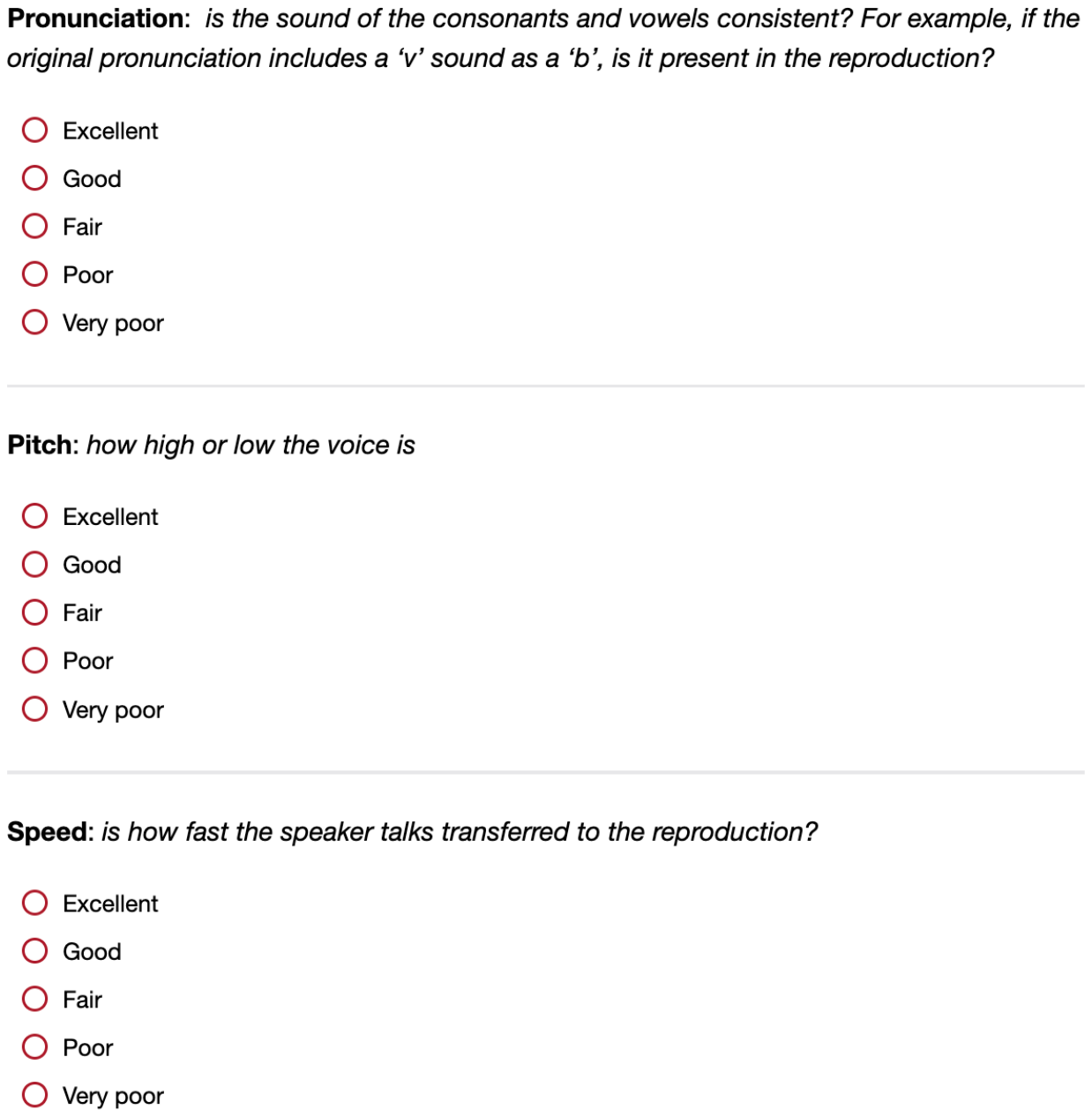}
    \caption{Sample survey participants completed, providing demographic information and evaluations of each cloned version of their voice. These evaluations followed the Accent Mimicry Questionnaire (AMQ) measure and detailed in \autoref{qual_ai_voice_clones}}.
    \label{fig:demo_AMQ}
\end{figure}

The following are the semi-structured interview questions for both in-person and remote sessions. See \autoref{fig:demo_AMQ} for the sample survey questions and \autoref{fig:interview_questions} for the semi-structured interview questions. Participants were assigned a sequential ID at the time the interview was conducted.

\vspace{-10mm}
\subsubsection{Demographic and Accent Mimicry Quality (AMQ) Questionnaires} \

\vspace{-40mm}
\begin{figure}[H]
    \centering
    \includegraphics[width=0.9\linewidth]{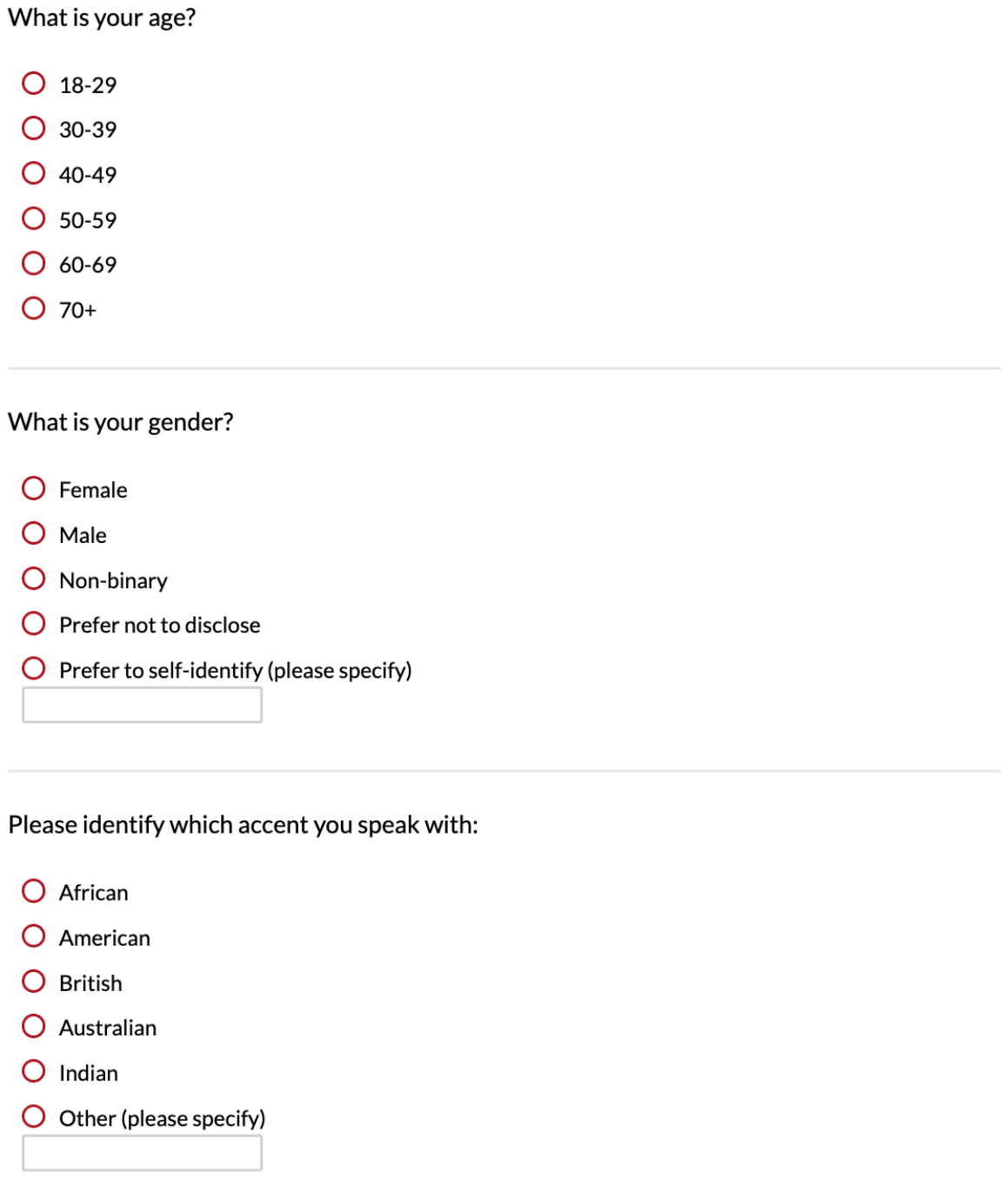}
    \label{fig:demo_AMQ_1}
\end{figure}

\vspace{-25mm}
\begin{figure}[H]
    \centering
    \includegraphics[width=0.9\linewidth]{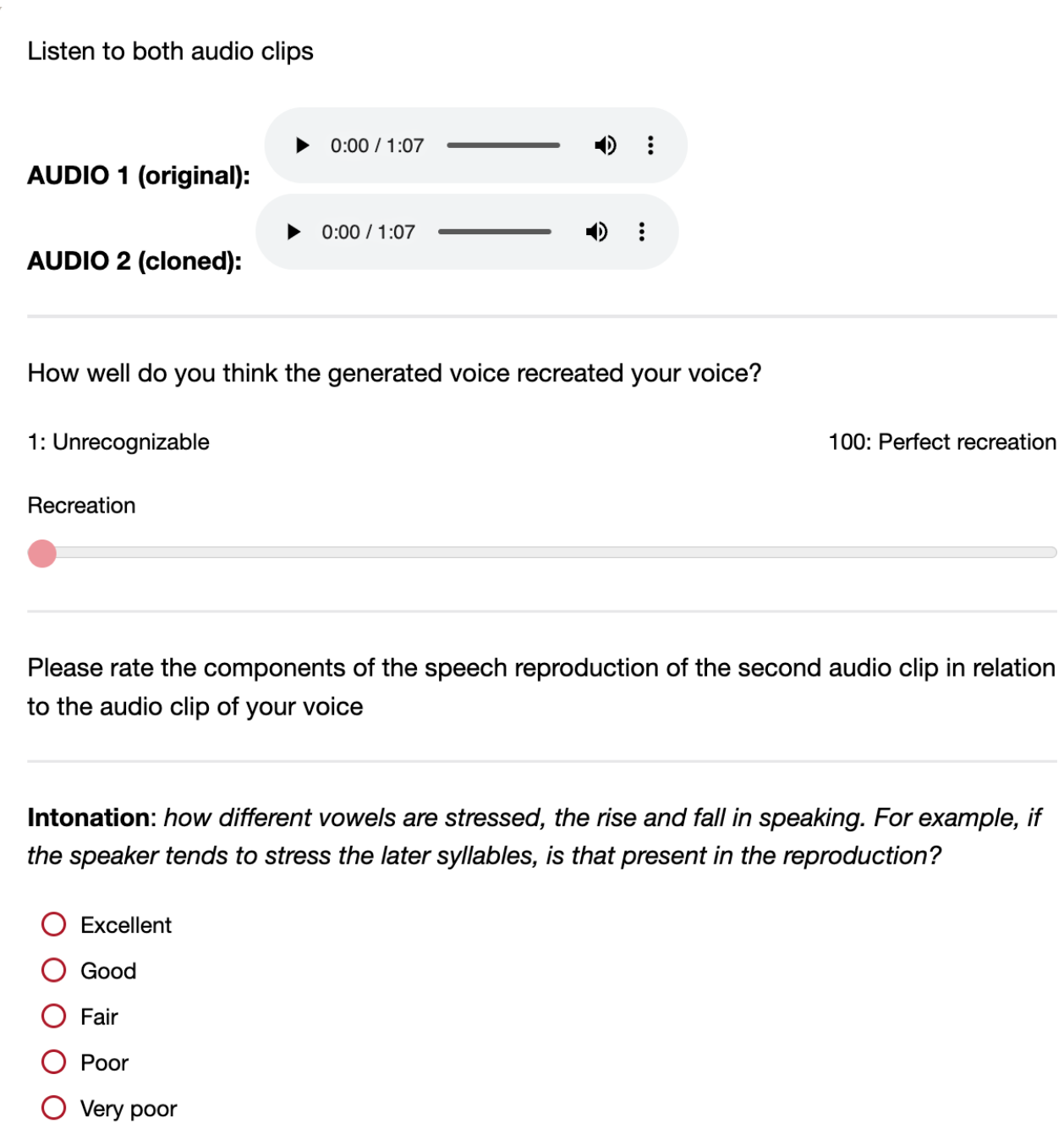}
    \label{fig:demo_AMQ_2}
\end{figure}

\vspace{-25mm}

\vspace{-5mm}

\begin{figure*}
\centering
\noindent\fbox{%
\parbox{0.95\textwidth}{%
\subsubsection{Voice and Accent Perception Questions}
\begin{itemize}
    \item What is your relationship with your voice and accent?
    \item What associations with your accent (e.g. positive, negative)?
    \item Do you have any feelings of insecurity or self-consciousness about your voice?
    \item Does your accent play a role culturally? At work? With friends?
    \item Do you feel that your accent is culturally significant?
    \item How does your accent influence your interactions in different contexts?
    \item Can you please share any experiences of how your accent shapes your identity and communication? Personality?    
\end{itemize}

\vspace{1em}
\subsubsection{AI Speech Technology Familiarity}
\begin{itemize}
    \item Are you familiar with AI speech generating services? 
    \item Have you heard of such services?
    \item Have you used any of these services before? If yes, how so?
    \item In what contexts do you think AI speech generating services are being used? Would you use it for any of those purposes?
\end{itemize}

\vspace{1em}
\subsubsection{Listening Experience Follow-up Questions (after each pair)}
\begin{itemize}
    \item How did you feel while listening to the generated voice clips (e.g. surprise, satisfaction, discomfort)?
    \item Do you feel like they represented you? 
    \item How accurate do you think the generated clips captured your accent and speech patterns (overall and per service)?
    \item In what way have the clips reflected your identity and communication style?
    \item Would you use a voice cloning service like this for podcasts or long-form audio content?
    \item What concerns do you have about using these services—authenticity, privacy, or ethical issues?
    \item What concerns do you have for others using these services?
\end{itemize}
}%
}
\caption{Semi-structured interview questions covering voice perception, AI familiarity, and reactions to generated voice samples.}
\label{fig:interview_questions}
\end{figure*}

\subsection{Intimate Relationship with Voice and Accent}
\label{sec:intimacy}
Our interviews of participants reflected how voice and accent characteristics have deep connections to feelings of representation, identity, and inclusivity, even before introducing them to the synthetic AI voice service experiments. We describe our findings below.

\subsubsection{\textbf{Experiences of Othering}}
Many participants recounted their experiences of being teased and alienated due to their voice and accent (P2, P4, P10, P12, P15, P17, P23, P24), as well as instances where their voice and accent were pointed out as markers of difference (P4, P6, P11, P13, P15, P24, P25, P26).
For participants like P4, P23, P10, and P25, their accents were linked to assumptions about their intelligence, with regional accents often perceived as indicators of lower intellect or lack of education. P4, with a Southern American accent, was particularly mindful of the negative connotations associated with it, recognizing how it could exacerbate stereotypes, especially as a Black woman. 
Similarly, P23, with a Northern British accent, noted how people in the U.K. assumed a lack of intelligence based on their accent, while people in the U.S. viewed a ``British twang'' as a sign of intelligence. 
Both P4 and P23 recalled how their parents taught them to modify their accents to combat these stereotypes and present themselves as more articulate.
P10, with an Indian accent, experienced assumptions about intelligence and career choices tied to the ``typical'' Indian accent, where people presumed they were smart in math or certain professions. 
In contrast, P25, with a Welsh accent, faced stereotypes of being ``slow'' or ``thick'', and consciously modified their accent to challenge those perceptions, often becoming more Welsh to make a point. 
P7, who struggled with the expectation to speak with a British or American accent, questioned why English had to be tied to specific accents, asking, ``\textit{...why am I supposed to adapt to these accents specifically? Why can't I just have my own accent?}''

Othering of voice and accent can be deeply isolating, as it leads individuals to feel disconnected in social and professional settings. For example, P12 described moments where they preferred not to speak at all because of a lack of confidence in their accent when speaking in a different language. Similarly, P14 shared how their Indian accent sometimes created barriers to communicating ideas, especially since certain sounds in English are absent in their native language. P14 expressed, ``\textit{I'm kind of scared...that I don't want to be misunderstood just because of my accent},'' highlighting the fear that a gap exists between what they intend to communicate and how it is perceived by others.
These experiences of othering, which contribute to social and cultural divides, align with prior research on accent-based discrimination and linguistic injustice~\cite{laurence2013english, piller2016linguistic, matsuda1991voices}.

\subsubsection{\textbf{Voice and Accent are Part of Identity}} 
For many participants, their voice and accent were viewed as an integral part of their personal and cultural identities.
50\% of participants expressed pride in their voice and accent because it represents who they are and where they come from. 
P11 confesses, ``\textit{I love the sound of my voice...I have an African accent because I was born and raised in Nigeria... I love [that] people have different accents based on their background and [it makes] me appreciate where I come from.}'' 
However, for some participants, this pride and acceptance of their voice and accent developed over time.
P9 reflects, 
\begin{quote}
    ``\textit{...I was probably a bit more self conscious about things, about myself than I probably needed to be. Um, just like...people don't like my voice or, like, people think my voice is weird...yeah, it probably caused me, like, a lot more, like, anxiety and depression growing up... It also kind of made me... [come to the idea to] accept yourself at some point in order to be happy.}''
\end{quote}
In contrast, a subset of participants with American accents reported their voice and accent was associated with little to no significance (P3, P8, P9, P12).
They viewed their voice and accent as not having strong emotional or social importance. 
These findings resonate with literature on social and self-identity, emphasizing how accents are integral to individual and group identity~\cite{edwards2009language}, serve as defining elements of social identity across various levels~\cite{joseph2004language}, and reflect broader dynamics of identity construction~\cite{bucholtz2005identity}.

\subsubsection{\textbf{Voice and Accent are Culture and Class Identifiers}}  
Relationship among voice, accent, and identity are multifaceted in identifying aspects of one's regional background, culture, and social class.
Among participants with British accents, accent was frequently highlighted as a marker of social class. 
P26 described how in the UK accents are ``an extremely important part of class'' and referred to the ``cultural quirk'' of being able to determine whether someone attended a private school based on just a few words. 
This ability to identify also reflects broader social narratives. 
P26 further shared that their accent is deeply rooted in their family history of coal mining and working-class origins, which they see as a source of pride and symbol of social mobility, even though it carries the weight of class divides that remain uncomfortable to discuss.
These findings connect with literature on how accents serve as identifiers which also reflect the subtle yet powerful role in shaping societal interactions~\cite{gluszek2010way, levon2022speaking, kraus2019evidence, dehghani2015subtlety}.

\subsubsection{\textbf{Code-switching as a Strategy}}
Code-switching emerged as a common strategy among participants across all accents. 
Many participants reported they code-switch depending on the context, the language spoken, the people around them, and the setting they are in. 
Participants also shared that they code-switched to help them be better understood, fit in, or accommodate others.
This adaptation can be seen as both a skill and a burden as described by P13: ``\textit{I think the whole reason I developed a different accent was just to fit in better to society. And so I think if I was to speak more like how I spoke at home, I wouldn't fit in as well.}'' On the other hand, some participants were resistant to change their voice and accent (P6, P10, P11, P18, P19, P20, P21, P22, P24, P25).
For example, P20 states, ``\textit{...I feel positive about my accent. And I'm like this is what I'm sticking to. I'm not going to try to change [for] anyone.}''
These insights align with literature and experiences shared on how people modify their speech and accent for social adaptation and accommodation~\cite{bell1984language, bucholtz2005identity, giles1991accommodation, Rangarajan2021}.

\end{document}